\documentclass[iop]{emulateapj}
\usepackage{amsmath,wasysym}
\usepackage{array,multirow,color,tablefootnote}
\usepackage[flushleft]{threeparttable}
\usepackage{float}
\usepackage{graphicx}
\usepackage{subfigure}
\usepackage{setspace}
\usepackage{epstopdf,textcomp}
\usepackage{natbib}

\bibpunct[ ]{(}{)}{;}{}{}{,}
\shorttitle{Transmission spectrum of HAT-P-12{\rm b}}
\shortauthors{Wong et~al.}

\begin{document}
\title{Optical to near-infrared transmission spectrum of the warm sub-Saturn HAT-P-12\MakeLowercase{b}}

\author{Ian Wong,\altaffilmark{1}\altaffilmark{$\dag$} Bj{\" o}rn Benneke,\altaffilmark{2} Peter Gao,\altaffilmark{3}\altaffilmark{$\dag$} Heather A. Knutson,\altaffilmark{4}  Yayaati Chachan,\altaffilmark{4} Gregory W. Henry,\altaffilmark{5} Drake Deming,\altaffilmark{6} Tiffany Kataria,\altaffilmark{7} Graham K.~H. Lee,\altaffilmark{8} Nikolay Nikolov,\altaffilmark{9,10,11} David K. Sing,\altaffilmark{12}  Gilda E. Ballester,\altaffilmark{13} Nathaniel J. Baskin,\altaffilmark{4} Hannah R. Wakeford,\altaffilmark{11} and Michael H. Williamson\altaffilmark{5}}
\affil{\textsuperscript{1}Department of Earth, Atmospheric, and Planetary Sciences, Massachusetts Institute of Technology, Cambridge, MA 02139, USA; iwong@mit.edu \\
   \textsuperscript{2}Department of Physics and Institute for Research on Exoplanets, Universit{\' e} de Montr{\' e}al, Montr{\' e}al, QC, Canada \\
   \textsuperscript{3}Department of Astronomy, University of California, Berkeley, CA 94720, USA \\
   \textsuperscript{4}Division of Geological and Planetary Sciences, California Institute of Technology, Pasadena, CA 91125, USA \\
   \textsuperscript{5}Center of Excellence in Information Systems, Tennessee State University, Nashville, TN 37209, USA \\
    \textsuperscript{6}Department of Astronomy, University of Maryland, College Park, MD 20742, USA \\
    \textsuperscript{7}Jet Propulsion Laboratory, California Institute of Technology, Pasadena, CA 91109, USA \\
    \textsuperscript{8}Atmospheric, Oceanic \& Planetary Physics, Department of Physics, University of Oxford, Oxford OX1 3PU, UK \\
    \textsuperscript{9}Astrophysics Group, School of Physics, University of Exeter, Exeter EX4 4QL, UK \\
    \textsuperscript{10}Department of Physics and Astronomy, Johns Hopkins University, Baltimore, MD 21218, USA \\
    \textsuperscript{11}Space Telescope Science Institute, 3700 San Martin Drive, Baltimore, MD 21218, USA \\
    \textsuperscript{12}Department of Earth and Planetary Sciences, Johns Hopkins University, Baltimore, MD 21218, USA \\
    \textsuperscript{13}Lunar and Planetary Laboratory, University of Arizona, Tucson, AZ 85721, USA\\
    \textsuperscript{$\dag$}\textit{51 Pegasi b} Fellow}

\begin{abstract}
We present the transmission spectrum of HAT-P-12b through a joint analysis of data obtained from the \textit{Hubble Space Telescope} Space Telescope Imaging Spectrograph (STIS) and Wide Field Camera 3 (WFC3) and Spitzer, covering the wavelength range 0.3--5.0~$\mu$m. We detect a muted water vapor absorption feature at 1.4~$\mu$m attenuated by clouds, as well as a Rayleigh scattering slope in the optical indicative of small particles. We interpret the transmission spectrum using both the state-of-the-art atmospheric retrieval code SCARLET and the aerosol microphysics model CARMA. These models indicate that the atmosphere of HAT-P-12b is consistent with a broad range of metallicities between several tens to a few hundred times solar, a roughly solar C/O ratio, and moderately efficient vertical mixing. Cloud models that include condensate clouds do not readily generate the sub-micron particles necessary to reproduce the observed Rayleigh scattering slope, while models that incorporate photochemical hazes composed of soot or tholins are able to match the full transmission spectrum. From a complementary analysis of secondary eclipses by Spitzer, we obtain measured depths of $0.042\%\pm0.013\%$ and $0.045\%\pm0.018\%$ at 3.6 and 4.5~$\mu$m, respectively, which are consistent with a blackbody temperature of $890^{+60}_{-70}$~K and indicate efficient day--night heat recirculation. HAT-P-12b joins the growing number of well-characterized warm planets that underscore the importance of clouds and hazes in our understanding of exoplanet atmospheres.
\end{abstract}

\section{Introduction}\label{sec:intro}
Over the past decade, major improvements in telescope capabilities and advancements in observation and analysis methods have enabled the intensive atmospheric characterization of an increasingly diverse population of exoplanets. Transmission spectroscopy has emerged as a powerful tool in studying the chemical composition of exoplanet atmospheres. By measuring the variations in transit depth as a function of wavelength, this technique directly probes the optically thin portion of the atmosphere along the day--night terminator of these tidally locked planets and is sensitive to various atmospheric components through their absorption signatures in the transmission spectrum. 

Transmission spectroscopy has hitherto successfully detected a broad range of chemical species in exoplanet atmospheres \citep[e.g.,][]{madhusudhan2014,demingseager}. Deriving estimates of the relative abundances of multiple atomic and molecular species yields constraints on more fundamental properties, such as disk-averaged metallicity, C/O ratio, and the temperature--pressure profile along the terminator. However, a large number of recent transmission spectroscopy studies have been confounded by the presence of clouds and hazes \citep[e.g.,][]{singstis}. Even trace amounts of cloud and haze particles can significantly increase the scattering opacity \citep[e.g.,][]{fortney2005,pont2008}, resulting in attenuation of absorption features in the transmission spectrum and reducing the ability to place meaningful constraints on key atmospheric properties, as, for example, in the cases of GJ 436b \citep{knutson2014} and GJ 1214b \citep{kreidberg}. Looking ahead to the future, a fuller understanding of the conditions under which clouds and hazes occur will be crucial in the selection of optimal targets with clear atmospheres for intensive observations using the limited time allocation available on next-generation telescopes, such as the James Webb Space Telescope \citep[JWST; e.g.,][]{bean2018,schlawin2018}.

As part of the continuing effort to better understand clouds in exoplanetary atmospheres, we examine in detail the transmission spectrum of HAT-P-12b. This planet is classified as a low-density sub-Saturn with a radius of 0.96~$R_{\mathrm{J}}$ and a mass of 0.21~$M_{\mathrm{J}}$, orbiting a K dwarf (0.73~$M_{\astrosun}$, 0.70~$R_{\astrosun}$, $T_{\mathrm{eff}}=4650$~K, [Fe/H]$ =-0.29$) with a period of 3.21~days \citep[][]{hartman}. Recent measurements of the Rossiter--McLaughlin effect for this system revealed a highly misaligned orbit \citep[$\lambda=-54^{+41}_{-13}$;][]{mancini}. A previous analysis showed that the near-infrared transmission spectrum was flat, indicating the presence of high-altitude aerosols \citep{line}. This planet has also been observed in transit at visible wavelengths both from the ground \citep{mallonn,alexoudi} and from space \citep{singstis,alexoudi}, with the latter studies revealing a slope in the optical transmission spectrum indicative of Rayleigh scattering by fine aerosol particles in the upper atmosphere \citep{barstow}.

The basic mechanisms for forming clouds and hazes on both solar system bodies and exoplanets involve either (a) condensation, in which a gaseous species changes phase to a liquid or solid upon becoming locally supersaturated either homogeneously or heterogeneously with the aid of condensation nuclei (e.g., \citealt{ackermanmarley2001}; \citealt{lodders2002}; \citealt{visscher2006}; \citealt{helling}; \citealt{visscher2010}; \citealt{charnay2018}; \citealt{gao}; \citealt{lee2018}; see also the reviews by \citealt{marley2013} and \citealt{helling2018}), or (b) photochemistry, induced by ultraviolet irradiation of the planet from the stellar host leading to the destruction of gaseous molecules and polymerization of the photolysis products into fine aerosol particles in the upper atmosphere \citep[e.g.,][]{zahnle2009,line2011,moses2011,venot2015,lavvas2017,horst,kawashima2018,adams2019}. Much of the detailed microphysics driving aerosol particle formation remains poorly understood, and models typically approximate haze formation using assumed chemical pathways, compositions, and formation efficiencies.



In addition to atmospheric metallicity, surface gravity, and the local temperature, secondary phenomena such as advection of material from the nightside to the dayside \citep[see, for example, the reviews by][]{showman2010,hengshowman2015}, the interplay between the degree of vertical mixing and particle size \citep[e.g.,][]{parmentier,zhangshowman}, and gravitational settling of particles \citep[e.g.,][]{lunine1989,marley1999,ackermanmarley2001,woitke2003,helling,gao2018} can affect the cloud properties at the terminator. The importance of clouds in interpreting and understanding exoplanet atmospheres has led to the development of increasingly complex cloud models incorporating many of the aforementioned chemical and physical processes \citep[e.g.,][]{helling2016,lee2016,lavvas2017,ohno2017,gao,kawashima2018,lines1,lines2,helling2019a,helling2019b,powell2019,woitke2019}.


Analyzing a planet's emission spectrum using secondary eclipse observations offers a complementary view of the atmosphere that may peer through the clouds that often obscure transmission spectra. This technique measures the outgoing flux from the planet's dayside hemisphere and provides independent constraints on dayside temperature, atmospheric metallicity, and cloud coverage. Both numerical models \citep[e.g.,][]{parmentier,lineparmentier,lines2,powell2018,caldas2019,helling2019a,helling2019b} and phase curve observations \citep[e.g.,][]{demory,shporerhu} suggest that clouds in exoplanet atmospheres are often localized to particular regions in the atmosphere, with incomplete coverage of the dayside hemisphere. In these instances, the planet's dayside emission spectrum is dominated by flux from the hotter, brighter cloud-free regions of the atmosphere and can yield additional insights into the atmosphere of planets with cloudy terminators, as in the case of HD 189733b \citep[][]{crouzet2014} and GJ 436b \citep[][]{morley2017}. 



In this paper, we analyze new near-infrared transit observations of HAT-P-12b obtained using the Wide Field Camera 3 (WFC3) instrument on the Hubble Space Telescope (HST) in spatial scan mode. Combining these data with previously published transit observations from the HST Space Telescope Imaging Spectrograph (STIS), WFC3, and Spitzer, we derive the transmission spectrum spanning the wavelength range 0.3--5.0~$\mu$m. Our analysis is supplemented by secondary eclipse measurements at 3.6 and 4.5~$\mu$m. In interpreting the results from our analysis, we utilize both atmospheric retrievals and predictions from microphysical cloud models to constrain the atmospheric properties of this planet. 

\section{Observations and Data Reduction}\label{sec:obs}
In this paper, we analyze a total of eight transit and four secondary eclipse observations obtained using three different instruments that span the wavelength range $0.3-5.0$~$\mu$m. This section provides a general overview of the methodology we use to extract light curves from the raw data for each of the three instruments. 

\subsection{HST WFC3}\label{subsec:wfc3}
As part of the Cycle 23 HST program GO-14260 (PI: D. Deming), we obtained time-resolved spectroscopic observations during two transits of HAT-P-12b on UT 2015 December 12 and 2016 August 31 using the G141 grism (1.0--1.7~$\mu$m) on WFC3. Each visit was comprised of five 96~minute HST orbits, with 45~minute gaps in data collection due to Earth occultations. The observations were carried out in spatial scan mode, with the star scanned perpendicularly to the dispersion direction across the detector at a rate of $0\overset{''}{.}03$~s$^{-1}$. In addition, at the start of the first orbit of each visit, we obtained an undispersed direct image of the star using the F139M grism for use in wavelength calibration. Each of the 74 spectra has a total exposure time of 112~s and extends roughly 30 pixels in the spatial direction. With the SPARS25 NSAMP=7 readout mode, each image file consists of seven nondestructive reads of the entire 266$\times$266 pixel subarray. These two scan mode visits have been previously analyzed in \citet{tsiaras2018}.

We also include in our analysis an older stare mode transit observation from UT 2011 May 29  (GO-12181; PI: D. Deming) that was analyzed previously in \citet{line} and \citet{singstis}. This visit consisted of 112 12.8~s exposures over the course of four orbits. Each orbit necessitated five buffer dumps, resulting in $\sim$9 minute gaps interrupting the data collection. There are 16 nondestructive reads of the 512$\times$512 pixel subarray in each image file. When reducing the images, we treat the stare mode data in the same way as the spatial scan mode observations. The observation details for the three WFC3 visits are summarized in Table~\ref{hubbleobservations}.

Starting with the dark- and bias-corrected \textit{*ima.fits} files produced by the standard WFC3 calibration pipeline, CALWFC3, we proceed with the data reduction using the Python 2--based Exoplanet Transits, Eclipses, and Phase Curves (ExoTEP) pipeline developed by B. Benneke and I. Wong \citep[see also][]{benneke2017, benneke2019}. To achieve maximal background subtraction in the extracted spectra, we follow a standard procedure for WFC3 spatial scan image processing \citep[e.g.,][]{deming2013,kreidberg,evans2016}: we construct subexposures by subtracting consecutive nondestructive reads and coadd all of the background-subtracted subexposures together to form the full background-corrected data frame. 

The spatial extent of each subexposure is determined by calculating the median flux profile for the difference image along the scan direction, i.e., $y$-direction, and locating the pixels where the flux falls to 20\% of the maximum value. To form the subexposure, we take the data that lie between these two rows, with an extra buffer of 15 pixels on the top and bottom, while setting all other pixel values to zero.  This method ensures that all of the stellar flux collected by the instrument between nondestructive reads is extracted and that the size of the extraction region for a given subexposure (e.g., difference of the third and second reads) remains largely consistent across each visit. The final results are not sensitive to the particular choice of buffer size between 10 and 20 pixels. The background level of each subexposure is set as the median of a 50 column wide region situated sufficiently far from the spectral trace and avoiding the edges of the subarray. 

Due to the particular geometry of the WFC3 instrument, the first-order spectrum of the G141 grism is not perfectly parallel to the detector rows. Also, there are significant variations in the length of the spectrum in the dispersion direction across the spatial scan, which results in the wavelength associated with a particular detector column varying from the top to the bottom. Lastly, imperfections in the pointing resets between each exposure lead to small horizontal shifts in the spectra across each visit \citep[e.g.,][]{deming2013, knutson2014, kreidberg}. Therefore, the shape of the spectrum on the detector is trapezoidal and slightly inclined relative to the subarray rows. Some previous analyses of WFC3 data have addressed this issue either by aligning the rows of the spectrum via interpolation \citep{kreidberg} or by deriving correction factors for the published wavelength calibration coefficients \citep{wilkins}.

In the ExoTEP pipeline, we follow the methodology described in detail in \citet{tsiaras} and compute the exact wavelength solution across the entire subarray for each exposure. In short, we first determine the position of the star along the $x$-axis of the detector for each exposure by taking the position of the star in the direct undispersed image, adjusting for differences in reference pixel location and subarray size between the direct and spatial scan images, and calculating the horizontal offset of each spectrum relative to the first spectroscopic exposure. The offsets are calculated by computing the centroid of each exposure and measuring the horizontal shift relative to the first exposure of the visit. 

Next, assuming that the spatial scan shifts the star position perfectly vertically across the detector, we determine the trace position and the wavelength solution along the trace using the calibration coefficients included in the configuration file \textit{WFC3.IR.G141.V2.5.conf} \citep{kuntschner} for a range of stellar $y$ positions. After a 2D cubic spline interpolation, we can now calculate the wavelength at every location on the subarray for each exposure. We also utilize this wavelength solution to apply a wavelength-dependent flat-field correction, using the cubic flat-field coefficients listed in the calibration file \textit{WFC3.IR.G141.flat.2.fits} \citep{kuntschner2,tsiaras}.

The last step in the ExoTEP data reduction process before light-curve extraction is cosmic-ray correction. For each exposure, we calculate the normalized row-added flux template. Next, we flag outliers using $5\sigma$ moving median filters of 10 pixels in width in both the $x$ and $y$ directions. Flagged pixel values are replaced by the value in the template corresponding to its $y$ position, appropriately scaled to match the total flux in its column. The particular parameters of the median filters are manually adjusted by inspecting the final corrected images and checking that all visible outliers have been removed. Due to the narrow vertical spatial profile of the trace in the stare mode images, we only apply the bad pixel correction in the horizontal direction for that visit.

To construct the spectroscopic light curves, we define a 20~nm wavelength grid from 1.10 to 1.66~$\mu$m and determine the spatial boundaries of the patch corresponding to each wavelength bin on the subarray using the previously derived wavelength solution. We calculate the flux within each patch by adding the pixel counts for all pixels that are fully within the patch and then computing the additional contribution from the partial pixels that are intersected by the patch boundaries. For each partial pixel, we integrate a local 2D cubic polynomial interpolation function over the subpixel regions that lie inside and outside of the given patch in order to compute the fraction of the total pixel count lying within the patch. This process ensures that the total flux is conserved and yields a modest reduction in the photometric scatter relative to more conventional extraction methods, which typically smooth the data in the dispersion direction prior to light-curve extraction \citep[e.g.,][]{deming2013,knutson2014,tsiaras}.

The time stamp for each data point is set to the mid-exposure time. To produce the broadband HST WFC3 light curve (i.e., white light curve), we simply sum the flux from the full set of individual spectroscopic light curves.

\begin{table}[t!]
	\centering
	\begin{threeparttable}
		\caption{HST Transit Observation Details} \label{hubbleobservations}	
		\renewcommand{\arraystretch}{1.2}
		\begin{center}
			\begin{tabular}{ l  m{0.1cm}  c m{0.1cm} c  c  c  } 
				\hline\hline
				Data Set & & UT Start Date & & $n_{\mathrm{exp}}$\textsuperscript{a} &  $t_{\mathrm{int}}$~(s)\textsuperscript{b}  & Mode  \\
				\hline
				WFC3 G141 & & & & & &\\
				\quad Visit 1 & & 2011 May 29 & & 112 & 12.8 & Stare \\
				\quad Visit 2 & & 2015 Dec 12 & & 74 & 112 & Scan \\
				\quad Visit 3 & & 2016 Aug 31 & & 74 & 112 & Scan \\
				STIS G430L & & & & & & \\
				\quad Visit 1 & & 2012 Apr 11 & & 34 & 280 & $\dots$ \\
				\quad Visit 2 & & 2012 Apr 30 &  & 34 & 280 & $\dots$ \\	
				STIS G750L & & & & & & \\
				\quad Visit 1 & & 2013 Feb 4 & & 34 & 280 & $\dots$ \\

				\hline

			\end{tabular}
			
			\begin{tablenotes}
				\small
				\item {\bf Notes.}
				\item \textsuperscript{a}Total number of exposures
				\item \textsuperscript{b}Total integration time per exposure
				
			\end{tablenotes}
		\end{center}
	\end{threeparttable}
\end{table}

\begin{table*}[t!]
	\centering
	\begin{threeparttable}
		\caption{Spitzer IRAC Observation and Data Reduction Details} \label{spitzerphotometry}	
		\renewcommand{\arraystretch}{1.2}
		\begin{center}
			\begin{tabular}{ l  m{0.1cm} c  m{0.1cm} c  c  c  c m{0.1cm} c  c  c } 
				\hline\hline
				Dataset & & UT Start Date & & $n_{\mathrm{img}}$\textsuperscript{a} &  $t_{\mathrm{int}}$~(s)\textsuperscript{b}  & $t_{\mathrm{trim}}$~(minutes)\textsuperscript{c}  & $r_{0}$\textsuperscript{c}  && $r_{1}$\textsuperscript{c}  & $r_{\mathrm{phot}}$\textsuperscript{c} & Binning\textsuperscript{d} \\
				\hline
				3.6~$\mu$m & && && & & && & & \\
				\quad Transit & &2013 Mar 8 && 8064 & 1.92 & 0 & 2.5 & & $\dots$  & 1.5 & 64 \\
				\quad Eclipse 1 && 2010 Mar 16 && 2097 & 10.4 & 60 & 3.0 && 2.0 & $\sqrt{\widetilde{\beta}}\times 1.3$& 16\\
				\quad Eclipse 2 && 2014 Apr 15 && 9024 & 1.92 & 30 & 4.0 && 1.0 & $\sqrt{\widetilde{\beta}}\times 1.5$& 16\\
				4.5~$\mu$m & &&& & & && & & & \\
				\quad Transit && 2013 Mar 11 && 8064 & 1.92 & 0 & 3.0 & & $\dots$ & 1.6 & 128 \\
				\quad Eclipse 1 && 2010 Mar 26 && 2097 & 10.4 & 45 & 3.5 && 2.5 & $\sqrt{\widetilde{\beta}}+0.7$ & 32 \\
				\quad Eclipse 2 && 2014 May 8 && 9024 &1.92 & 60 & 2.5 && $\dots$ & 2.4 & 128 \\

				\hline

			\end{tabular}
			
			\begin{tablenotes}
				\small
				\item {\bf Notes.}
				\item \textsuperscript{a}Total number of images.
				\item \textsuperscript{b}Total integration time per image.
				\item \textsuperscript{c}Here $t_{\mathrm{trim}}$ is the amount of time trimmed from the start of each time series prior to fitting, $r_{0}$ is the radius of the aperture used to determine the star centroid position, and $r_{1}$ is the radius of the aperture used to compute the noise pixel parameter $\widetilde{\beta}$. The $r_{\mathrm{phot}}$ column denotes how the photometric extraction aperture is defined. All radii are given in units of pixels. When using a fixed aperture, the noise pixel parameter is not needed, so $r_{1}$ is undefined. See text for more details.
				\item \textsuperscript{d}Number of data points placed in each bin when binning the photometric series prior to fitting.
				
			\end{tablenotes}
		\end{center}
	\end{threeparttable}
\end{table*}

\subsection{HST STIS}\label{subsec:stis}
We observed three transits of HAT-P-12b with the HST STIS instrument as part of the program GO-12473 (PI: D. Sing). Observations of two transits were carried out using the G430L grating (290--570~nm) on UT  2012 April 11 and 30; the third transit was observed using the G750L grating (550--1020~nm) on UT 2013 February 4. The two gratings used have resolutions of $R=530-1040$ (5.5 and 9.8~\AA~per 2 pixel resolution element for the G430L and G750L gratings, respectively). Each visit contains a total of 34 science exposures across four HST orbits, with the third orbit occurring during mid-transit. To reduce overhead, data were read out from a 1024$\times$128 subarray with a per-exposure integration time of 280~s. The observational details for the three STIS visits are listed in Table~\ref{hubbleobservations}. This set of observations has been analyzed in two previous independent studies: \citet{singstis} and \citet{alexoudi}.

The raw images are flat-fielded using the latest version of CALSTIS. The subsequent data reduction is completed using the ExoTEP pipeline. We remove outlier pixel values in the time series by first computing the median image across each visit and then replacing all pixel values in the individual exposure frames varying by more than $4\sigma$ with the corresponding value in the median image. We apply the wavelength solution  provided in the \textit{*sx1.fits} calibrated files and extract the column-added 1D spectra, choosing the aperture width and whether to subtract the background so as to minimize the scatter in the residuals from the transit light-curve fit \citep[e.g.,][]{deming2013}. In our analysis of the two G430L observations, we utilize 9 and 7 pixel wide apertures, respectively, removing the background for the first visit only; in the case of the G750L transit, we find that extracting spectra from a 7 pixel wide aperture after background subtraction results in the minimum scatter.

Data collected using the G750L grism suffer from a fringing effect, which manifests itself as an interference pattern superposed on the 1D spectrum and is especially apparent at wavelengths longer than 700~nm. Following the methods outlined in previously published analyses of data from this program \citep[e.g.,][]{nikolov,nikolov2,singstis}, we defringe our data using a fringe flat frame obtained at the end of the G750L science observations. 

Lastly, we correct for subpixel wavelength shifts in the dispersion direction across each visit by fitting for the horizontal offsets and amplitude scaling factors that align all extracted spectra with the first one. The normalized broadband light curve is simply the time series of the optimized amplitude scaling factors. To generate the spectroscopic light curves, we collect the flux within 200 and 100 pixel bins for the G430L and G750L observations, respectively. The wavelength bounds corresponding to the 200 pixel bins for the two G430L transit observations differ by less than the characteristic wavelength resolution element (0.55 nm). For the G750L dataset, we also include two narrow wavelength bins centered around the sodium and potassium absorption lines (588.7--591.2 and 770.3--772.3~nm respectively).

\subsection{Spitzer IRAC}\label{subsec:spitzer}

Two transits of HAT-P-12b were observed in the 3.6 and 4.5~$\mu$m broadband channels of the Infrared Array Camera (IRAC) on the Spitzer Space Telescope (Program ID 90092; PI: J.-M. D{\' e}sert). The observations took place on UT 2013 March 8 and 11 and were carried out in subarray mode, which produces 32$\times$32 pixel ($39''\times39''$) images centered on the stellar target. Each transit observation is comprised of 8064 images with a per-exposure effective integration time of 1.92~s.

A set of two secondary eclipse observations, one in each of the two postcryogenic IRAC channels, was obtained on UT 2010 March 16 and 26 (Program ID 60021; PI: H. Knutson). These data consist of 2097 images per passband obtained in full array mode at a resolution of 256$\times$256 pixels ($5\overset{''}{.}2\times5\overset{''}{.}2$) with an effective exposure time of 10.4~s per image. Peak-up pointing was utilized, which entails an initial 30 minute observation prior to the start of the science observation to allow for the stabilization of the telescope pointing. These eclipses were previously analyzed in \citet{todorov}. A second set of hitherto unpublished secondary eclipse observations, including one in each channel, was obtained on UT 2014 April 15 and May 8 (Program ID 10054; PI: H. Knutson). These observations were taken in subarray mode with peak-up pointing and contain 9024 images with effective exposure times of 1.92~s.

We extract photometry following the techniques described in detail in previous analyses of postcryogenic Spitzer data \citep[e.g.,][]{knutson2012,lewis,todorov,wong2,wong3}. Starting with the dark-subtracted, flat-fielded, linearized, and flux-calibrated images produced by the standard IRAC pipeline, we calculate the sky background via a Gaussian fit to the distribution of pixel values, excluding pixels near the star and its diffraction spikes, as well as the problematic top (32nd) row, which has flux values that are systematically lower than the other rows. We also iteratively trim outlier pixel values on a pixel-by-pixel basis using a $3\sigma$ moving median filter across the adjacent 64 images in the time series.

The position of the star on the detector is determined using the flux-weighted centroiding method \citep[e.g.,][]{knutson2012}. The width of the star's point response function (PRF; i.e., the convolution of the star's point-spread function and the detector response function) is estimated by computing the noise pixel parameter $\widetilde{\beta}$ \citep[see][for a full discussion]{lewis}. The stellar position and PRF width are calculated using circular apertures of radius $r_{0}$ and $r_{1}$, respectively, which we vary in 0.5~pixel steps to produce different versions of the extracted photometry. The photometric series can be extracted using both fixed and time-varying circular apertures, where in the case of time-varying apertures, the radii are related to the square root of the noise pixel parameter by either a constant scaling factor or a constant shift \citep[e.g.,][]{wong2,wong3}.

Prior to fitting (see Section~\ref{sec:analysis}), we can bin the photometric series into various intervals equal to powers of two (i.e., 1, 2, 4, 8, etc. points). To aid in the removal of instrumental systematics, we also experiment with trimming the first 15, 30, 45, or 60 minutes of data from the time series. Before fitting each photometric series with our transit/eclipse light-curve model, we apply an iterative moving median filter of 64 data points in width to remove points with measured fluxes, $x$ or $y$ star centroid positions, or $\sqrt{\widetilde{\beta}}$ values that vary by more than 3$\sigma$ from the corresponding median values. For all Spitzer datasets, the number of removed points is less than 5\% of the total number of data points, and slightly altering the width of the median filter does not significantly affect the number of removed points.

For each $Spitzer$ transit or secondary eclipse observation, we determine the optimal aperture and photometric parameters by fitting the various photometric series with the model light curve and selecting the version that minimizes the scatter in the resultant residuals, binned in 5 minute intervals \citep{wong2,wong3}. The optimal values are listed in Table~\ref{spitzerphotometry}. 

\subsection{Photometric monitoring for stellar activity}\label{subsec:monitoring}

High levels of chromospheric activity, which can lead to significant photometric variability and incur wavelength-dependent  biases in the measured transmission spectrum \citep[e.g.,][]{rackham}, can be displayed by K dwarfs such as HAT-P-12. In particular, the presence of unocculted starspots can impart slope changes to the shape of the transmission spectrum in the optical, affecting the interpretation of the planet's atmospheric  properties.

To characterize the level of stellar activity on HAT-P-12, we obtained Cousins $R$-band photometry of the star using the the Tennessee State University Celestron 14 inch (C14) Automated
Imaging Telescope (AIT) located at Fairborn Observatory, Arizona. Differential magnitudes of HAT-P-12 were calculated relative to the mean brightnesses of five constant comparison stars from five to ten co-added consecutive exposures. Details of our observing, data reduction, and analysis procedures with the AIT are described in \citet{sing2015}.

\begin{figure}[t!]
\begin{center}
\includegraphics[width=\linewidth]{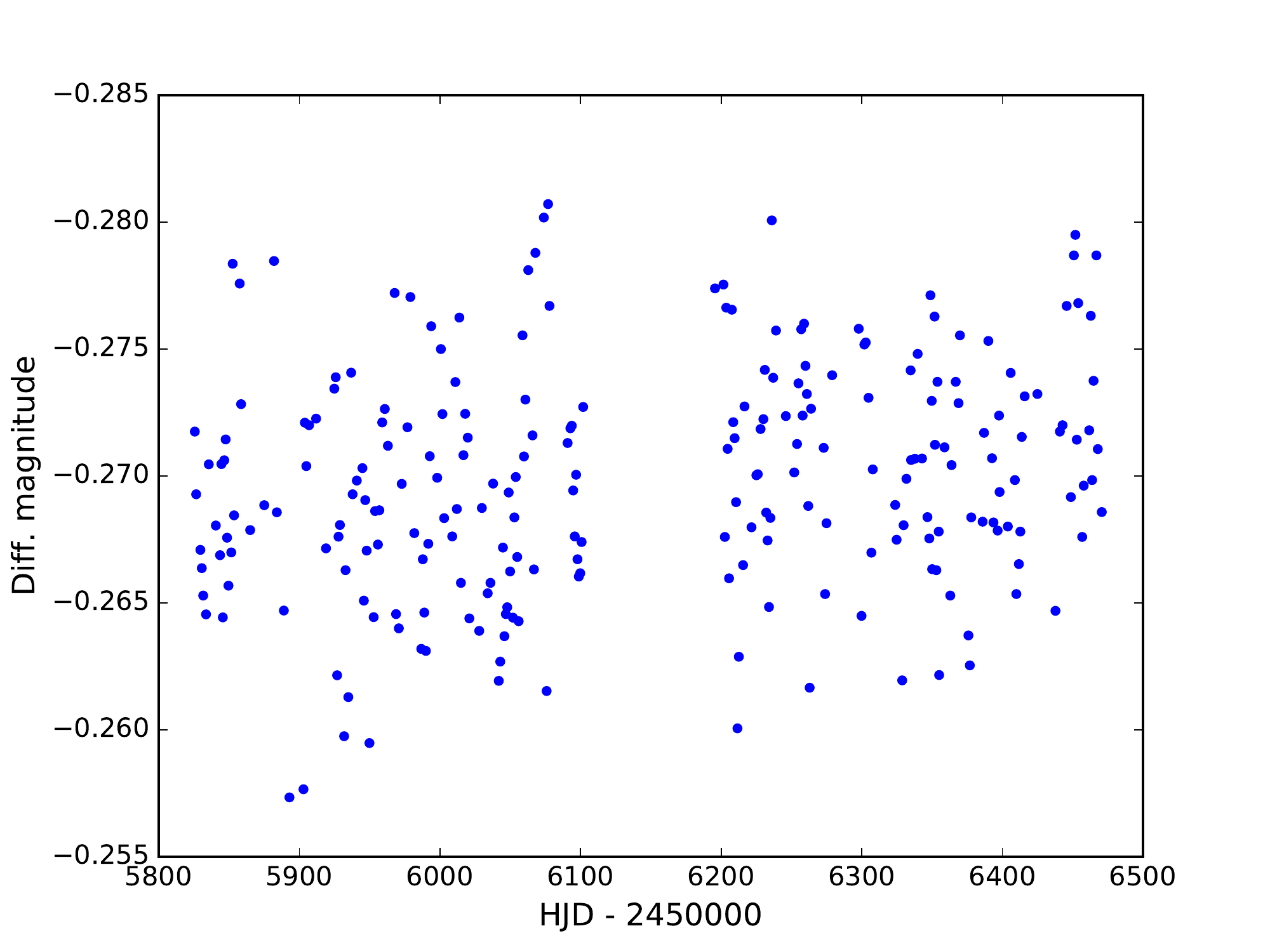}
\end{center}
\caption{Composite $R$-band nightly differential photometry of HAT-P-12 for the 2011--2012 and 2012--2013 observing seasons, obtained with the C14 AIT at Fairborn Observatory. The standard deviation of the data is 0.0046 and 0.0042 for the two seasons, comparable to the measurement precision, indicating that HAT-P-12 shows no significant variability. } \label{photovar}
\end{figure}

A total of 237 successful nightly observations were collected across two observing seasons (season 1: 2011 September 20--2012 June 22; season 2: 2012 September 24--2013 June 26). The individual observations are plotted in Figure~\ref{photovar}. The seasonal means in differential magnitude are $-0.2689\pm0.0004$ and $-0.2708\pm0.0004$, with corresponding single observation standard deviations of 0.0046 and 0.0042, respectively. These scatter values are comparable to the approximate limit of the measurement precision. This indicates that HAT-P-12 does not show any significant variability.

When performing a periodogram analysis of individual seasonal datasets, we do not retrieve any frequencies that produce amplitudes larger than the seasonal standard deviations. In particular, we do not detect a variability signal with a period near the estimated rotational period of the star \citep[$P_{\mathrm{rot}} \sim44$~days;][]{mancini}. Such a periodicity in the photometry would be indicative of rotational modulation of weak features on the stellar surface.
We therefore conclude that HAT-P-12 is a very quiescent host star.

This conclusion is consistent with the findings from high-resolution spectroscopy of the host star during the initial discovery and characterization of the system \citep{hartman}, which did not detect significant levels of variability suggestive of large starspots across the stellar surface. Analyses of stellar spectra from the Keck High Resolution Echelle Spectrometer (HIRES) and more recently from the High Accuracy Radial Velocity Planet Searcher (HARPS) instrument also indicate low stellar activity as determined by the Ca II H and K lines: $\log(R'_{\mathrm{HK}})=-5.104$ \citep{knutson2010} and $\log(R'_{\mathrm{HK}})=-4.9$ \citep{mancini}. In addition, none of the transit light curves analyzed in this work show evidence for occulted spots. 

It is notable that HAT-P-12b has an optical transmission spectrum that shows a slope indicative of Rayleigh scattering \citep[][and this work]{singstis,alexoudi}, while the host star has low stellar activity. This is in contrast to the paradigmatic case of HD 189733b, which has a clear optical scattering slope and an active host star. Thus, HAT-P-12b serves as an important test of whether the transmission slope is related to stellar activity, which could happen in the case of unocculted stellar spots or with enhanced photochemistry as a product of higher stellar far- and near-UV levels.

\section{Analysis}\label{sec:analysis}
We carry out a global analysis of all eight transit light curves (three HST WFC3 G141 visits, two HST STIS G430L visits, one HST STIS G750L visit, and two Spitzer IRAC visits at 3.6 and 4.5~$\mu$m) by simultaneously fitting our transit light-curve model, instrumental systematics models, and photometric noise parameters using the ExoTEP pipeline \citep{benneke2017, benneke2019}. We also perform an independent combined fit of the four Spitzer secondary eclipse light curves.

\subsection{Broadband Light-Curve Fits}\label{subsec:wlc}

\subsubsection{Instrumental Systematics}\label{systematics}
Prior to fitting the HST WFC3 light curves, we discard the first orbit, as well as the first two exposures of each orbit, which notably improves the resultant fits. We also remove the 31st and 68th exposures, which were affected by cosmic ray hits, from the second spatial scan mode transit light curve. 

Raw uncorrected light curves obtained using the HST WFC3 instrument exhibit well-documented systematic flux variations across the visit, as well as within each individual spacecraft orbit \citep[e.g.,][]{deming2013,kreidberg}.  We model the HST WFC3 instrumental systematics with the following analytical function \citep[e.g.,][]{berta,kreidberg2015}:
\begin{equation}\label{wfc3systematics}S_{\mathrm{WFC3}}(t) = (c+vt_v)\cdot(1-\exp[-at_{\mathrm{orb}}-b-D(t)]).\end{equation}
Here $c$ is a normalization constant, $v$ is the visit-long slope, $a$ and $b$ are the rate constant and amplitude of the orbit-long exponential ramps, and $t_{v}$ and $t_{\mathrm{orb}}$ are the time elapsed since the beginning of the visit and since the beginning of the orbit, respectively. Here $D(t)$ is set to a constant $d$ for points in the first fitted orbit and zero everywhere else, reflecting the observed difference in the ramp amplitude between the first fitted orbit and the subsequent orbits.

We find that stare mode observations exhibit an additional quasi-linear systematic trend across exposures taken between each buffer dump (five per orbit for our visit). We can correct for this trend by appending an extra factor of $(1+dt_{d})$ to Eq.~\eqref{wfc3systematics}, where $d$ is the linear slope, and $t_{d}$ is the time elapsed since the end of the last buffer dump. 

Similar ramp-like instrumental systematics are also apparent in HST STIS raw light curves, albeit with a somewhat different shape. We correct these systematics using a standard analytical model \citep{sing2008},
\begin{equation}\label{stissystematics}S_{\mathrm{STIS}}(t) = (c+vt_v)\cdot(1+p_1 t_{\mathrm{orb}}+p_2 t_{\mathrm{orb}}^2+p_3 t_{\mathrm{orb}}^3+p_4 t_{\mathrm{orb}}^4),\end{equation}
where $c$, $v$, $t_{v}$, and $t_{\mathrm{orb}}$ are defined in the same way as in Eq.~\eqref{wfc3systematics}, and the coefficients $p_{1-4}$ describe the fourth-order polynomial shape of the orbit-long trend. As with the HST WFC3 light curves, we remove the first orbit, as well as the first two exposures of each orbit prior to fitting.

Raw photometry obtained using the Spitzer IRAC instrument is characterized by short-timescale variations in the measured flux due to small oscillations of the telescope pointing and nonuniform sensitivity of the detector at the subpixel scale. We correct for these intrapixel sensitivity variations by using the modified version of the Pixel Level Decorrelation method \citep[PLD;][]{pld} described in \citet{benneke2017}:
\begin{equation}\label{pld}S_{\mathrm{IRAC}}(t) = 1+\sum\limits_{k=1}^{9}w_{k}\hat{P}_{k}(t_i)+ vt_i.\end{equation}

The arrays $\hat{P}_{k}$ represent the pixel counts for the nine pixels located in a $3\times 3$ box centered on the star's centroid position normalized to sum to unity at each point in the time series. These normalized pixel count arrays are placed into a linear combination with weights $w_{k}$. The last term models a visit-long linear trend, where $v$  is the slope parameter and $t_{i}$ denotes the time elapsed since the beginning of the time series. As with the photometric series, the pixel count arrays can be binned prior to fitting. We optimize for the binning interval and the number of points trimmed from the start of the observation by carrying out individual fits of each IRAC transit light curve (see Section~\ref{subsec:spitzer}). In the global transit light-curve fit, no additional alterations of the IRAC light curves are needed.

\begin{table*}[t!]
\small
	\centering
	\begin{threeparttable}
		\caption{Global Broadband Light-curve Fit Results} \label{tab:fit}	
		\renewcommand{\arraystretch}{1.2}
		\begin{center}
			\begin{tabular}{ l c c c} 
				\hline\hline
				Parameter & Instrument & Wavelength (nm) & Value \\
				 \hline
				Planet radius, $R_{p}/R_{*}$ & STIS G430L & 289--570 &  $0.13798\pm0.00069$ \\
				Planet radius, $R_{p}/R_{*}$ & STIS G750L & 526--1025 & $0.13915^{+0.00053}_{-0.00054}$ \\
				Planet radius, $R_{p}/R_{*}$ & WFC3 G141 & 920--1800 &  $0.13743^{+0.00017}_{-0.00016}$\\
				Planet radius, $R_{p}/R_{*}$ & IRAC 3.6~$\mu$m & 3161--3928 & $0.13627^{+0.00074}_{-0.00068}$ \\
				Planet radius, $R_{p}/R_{*}$ &  IRAC 4.5~$\mu$m & 3974--5020 &  $0.1386^{+0.0014}_{-0.0015} $\\
				Transit center time, $T_{0}$ (BJD$_\mathrm{TDB}$) & $\dots$ & $\dots$ & $2,357,368.783203\pm0.000025$\\
				Period, $P$ (days) & $\dots$ & $\dots$ & $3.21305831\pm0.00000024$ \\
				Impact parameter, $b$ & $\dots$ & $\dots$ & $0.272^{+0.016}_{-0.017}$ \\
				Inclination,\textsuperscript{a} $i$ (deg) & $\dots$ & $\dots$ &  $88.655^{+0.090}_{-0.084}$\\
				Relative semimajor axis, $a/R_{*}$ & $\dots$ & $\dots$ &$11.574^{+0.055}_{-0.054}$ \\
				
				\hline
			\end{tabular}
			
			\begin{tablenotes}
			 \item {\bf Note.} \textsuperscript{a}Inclination derived from impact parameter via $b=(a/R_{*})\cos i$.
			\end{tablenotes}
		\end{center}
	\end{threeparttable}
\end{table*}

\begin{figure*}[h!]
\begin{center}
\includegraphics[width=12cm]{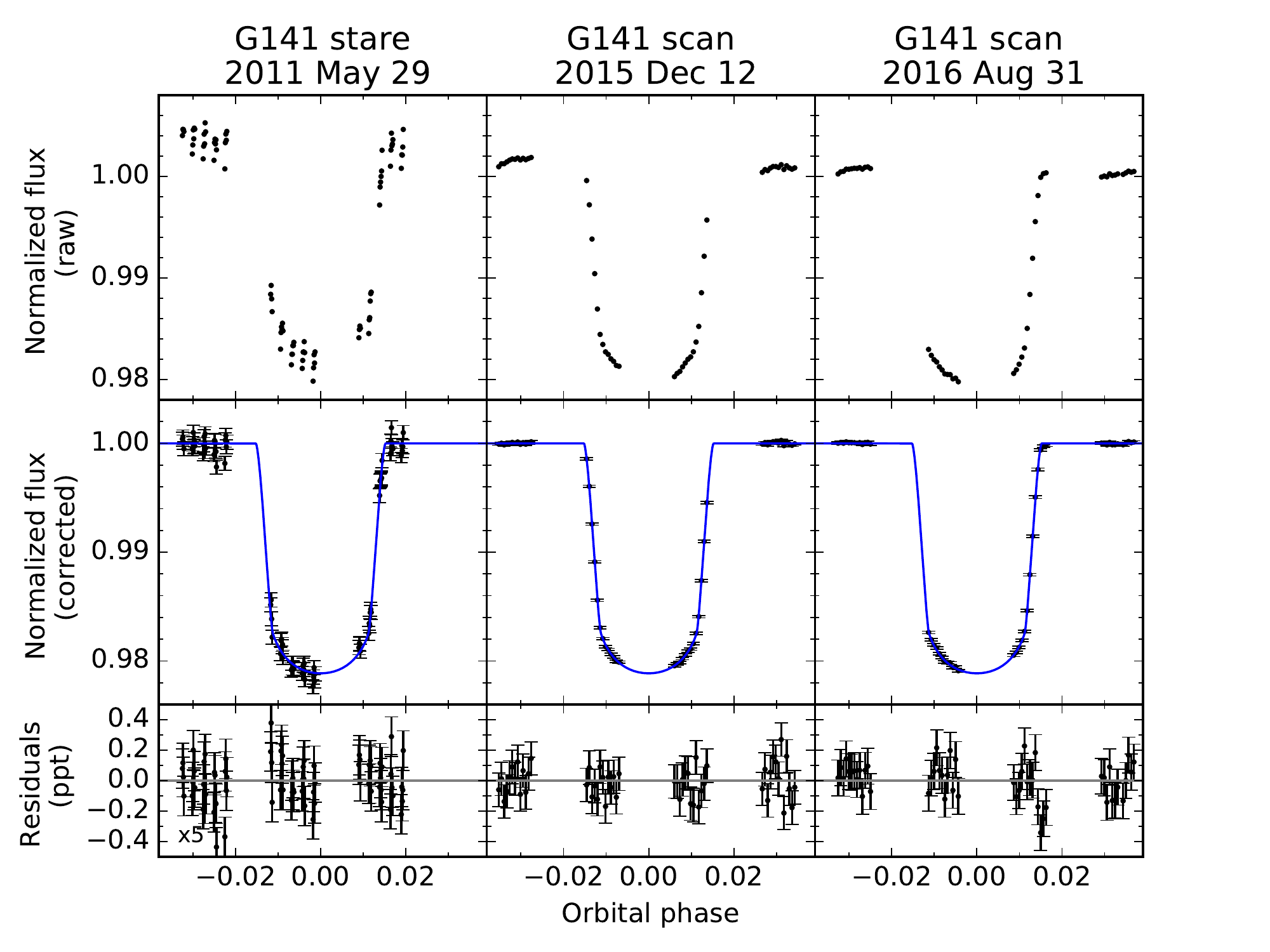}
\end{center}
\caption{Raw (top) and instrument systematics-corrected (middle) broadband curves of the three transits observed using the HST WFC3 G141 grism (1.1--1.7~$\mu$m). The best-fit transit light curve is shown in blue. The bottom panels show the resulting residuals after removing the best-fit instrumental model and transit light curve. The error bars on each data point have been set to the best-fit photometric noise parameter. Note that the residuals plotted for the stare mode transit have been divided by a factor of 5 in order to display them using the same y-axis scale.} \label{wfcwlc}
\end{figure*}

\subsubsection{Limb Darkening}

The ExoTEP pipeline incorporates the Python-based package LDTK \citep{ldtk} to automatically calculate limb-darkening coefficients. Given the literature values and uncertainties for the stellar parameters ($T_{\mathrm{eff}}=4650\pm60$~K, $\log g=4.61\pm0.01$, $\mathrm{[Fe/H]}=-0.29\pm0.05$; \citealt{hartman}), this program generates a mean limb-darkening profile and profile uncertainties for each specified bandpass (broadband or spectroscopic) via Monte Carlo sampling of interpolated $50-2600$~nm PHOENIX stellar intensity spectra \citep{husser} within a $3\sigma$ range in the space of the three stellar parameters. 

Subsequent maximum-likelihood optimization returns the best-fit linear, quadratic, or nonlinear limb-darkening coefficients to be used in calculating the transit shape in each bandpass. In our global fit, we find that using the four-parameter nonlinear limb-darkening model yields the lowest residual scatter during ingress and egress, particularly for the high signal-to-noise HST WFC3 spatial scan mode visits.

Because the custom stellar spectra accessed by the LDTK package do not cover wavelengths longer than 2.6~$\mu$m, we set the limb-darkening coefficients for the Spitzer IRAC 3.6 and 4.5~$\mu$m transit light curves to the values computed following the methods described in \citet{sing}. These coefficients are tabulated online\footnote{\texttt{pages.jh.edu/$\sim$dsing3/David\_Sing/Limb\_Darkening.html}} for a wide range of ($T_{\mathrm{eff}}$, $\log g$, $z$) values, and we choose the values listed for the set of stellar parameters closest to the literature values for HAT-P-12.

To empirically verify that our choice of fixing limb-darkening coefficients to modeled or tabulated values does not have a significant effect on the measured transmission spectrum, we have experimented with fitting for quadratic limb-darkening coefficients in individual fits of the WFC3 scan mode visits and the broadband Spitzer transit light curves; these visits have either complete transit coverage or the highest per-point precision. We find that the fitted coefficients have large relative uncertainties (20\%--70\%), i.e., are not well constrained by the data, while being statistically consistent with the corresponding values produced by LDTK or listed in the \citet{sing} tables. Crucially, no significant shifts in transit depth occur when switching from fixed limb-darkening coefficients to fitted values.

\begin{figure*}[h!]
\begin{center}
\includegraphics[width=12cm]{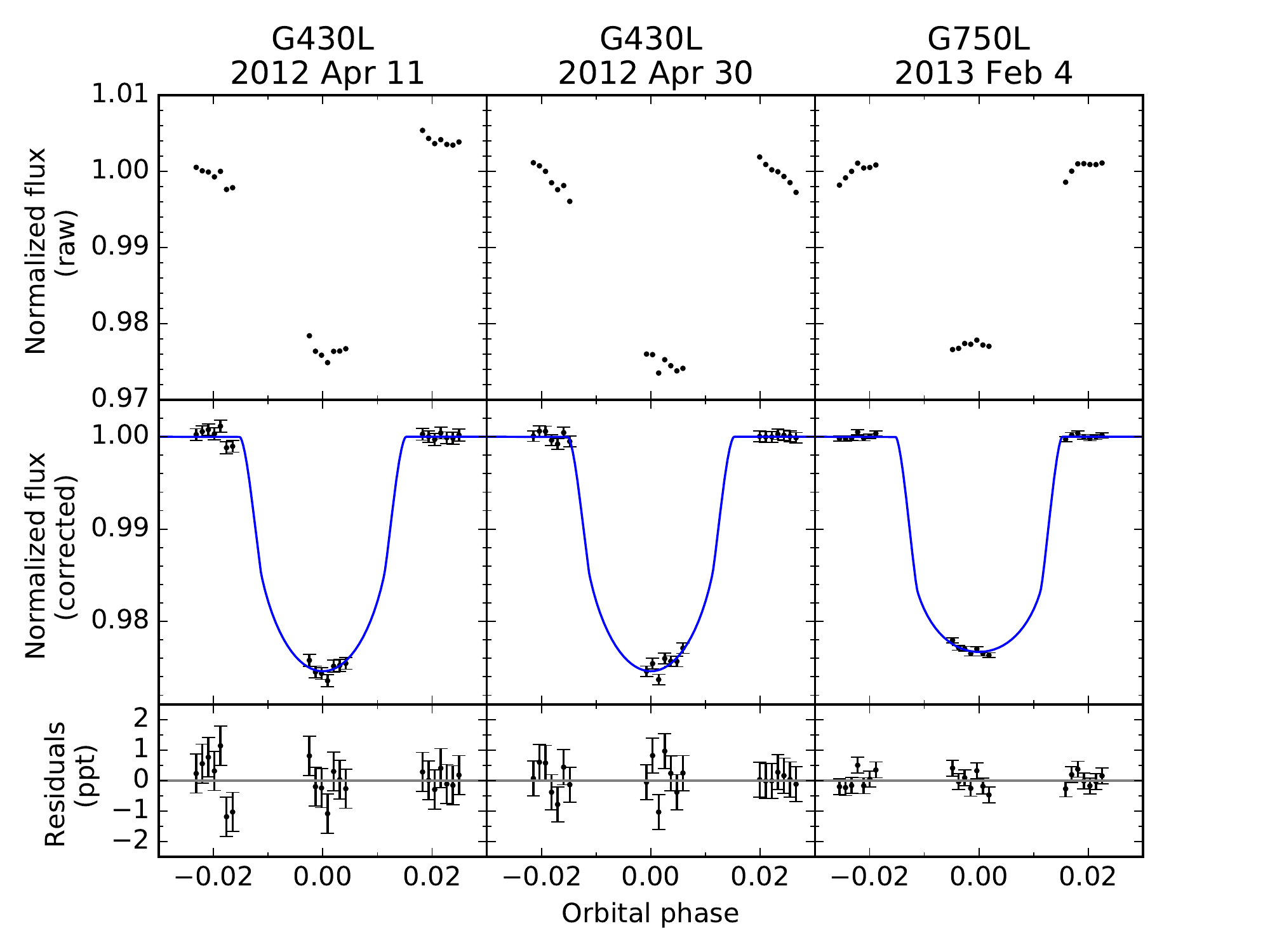}
\end{center}
\caption{Same as Figure~\ref{wfcwlc}, but for the three transit observations obtained using the HST STIS G430L and G750L grisms.} \label{stiswlc}
\end{figure*}

\begin{figure*}[t!]
\begin{center}
\includegraphics[width=12cm]{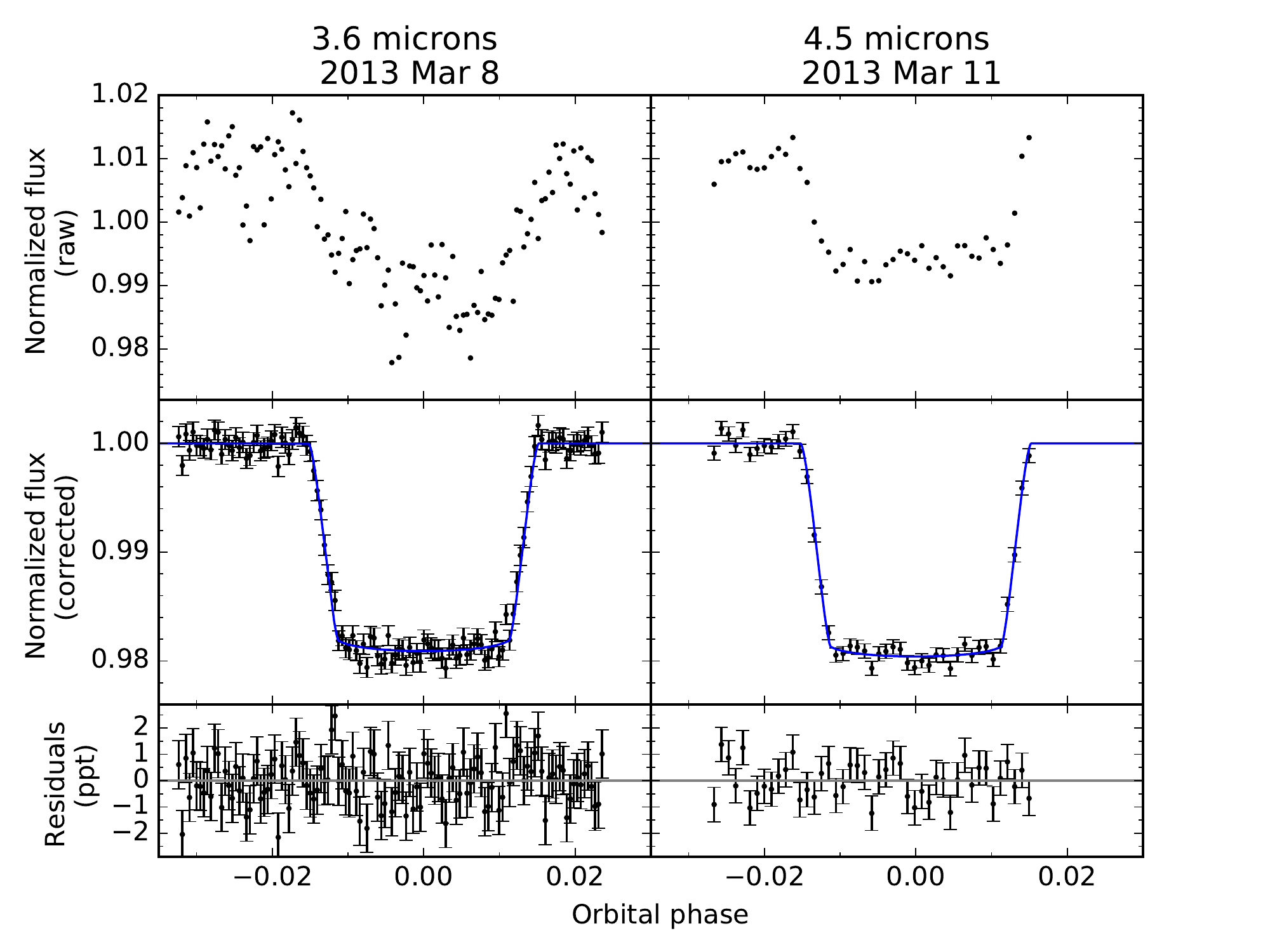}
\end{center}
\caption{Same as Figure~\ref{wfcwlc}, but for the 3.6~$\mu$m and 4.5~$\mu$m Spitzer IRAC transit datasets. The data are shown binned into 64- and 128-point bins, respectively, as was done prior to the global broadband transit light-curve fit.} \label{spitzerwlc}
\end{figure*}

\subsubsection{Global Fit Results}

\begin{table}[t!]
\small
	\centering
	\begin{threeparttable}
		\caption{Spectroscopic Light-curve Fit Results} \label{tab:fit2}	
		\renewcommand{\arraystretch}{1.2}
		\begin{center}
			\begin{tabular}{ l m{1cm} c} 
				\hline\hline
				Wavelength (nm) & & $R_{p}/R_{*}$ \\
				 \hline
				 STIS G430L & & \\
				 346--401 & & $0.1418\pm0.0028$\\
				 401--456 & & $0.1405\pm0.0012$\\
				 456--511 & & $0.1387\pm0.0008$\\
				 511--565 & & $0.1390\pm0.0008$\\
				  & & \\
				 STIS G750L & & \\
				 528--577 & & $0.1392\pm0.0011$\\
				 577--626 & & $0.1378\pm0.0013$\\			 
				 626--675 & & $0.1394\pm0.0009$\\
				 675--723 & & $0.1377\pm0.0008$\\
				 723--772 & & $0.1388\pm0.0007$\\
				 772--821 & & $0.1386\pm0.0019$\\
				 821--870 & & $0.1379\pm0.0013$\\
				 870--919 & & $0.1364\pm0.0015$\\
				 919--968 & & $0.1370\pm0.0021$\\
				 968--1016 & & $0.1369\pm0.0029$\\ 
				 588.7--591.2 (Na)\textsuperscript{a} & & $0.1357\pm0.0032$\\
				 770.3--772.3(K)\textsuperscript{a} & & $0.1391\pm0.0052$\\
				  & & \\
				 WFC3  G141& & \\
				1100--1120 & & $0.13666\pm0.00050$ \\
				1120--1140 & & $0.13834\pm0.00047$\\
				1140--1160 & & $0.13794\pm0.00045$\\
				1160--1180 & & $0.13744\pm0.00042$\\
				1180--1200 & & $0.13682\pm0.00036$\\
				1200--1220 & & $0.13779\pm 0.00044$\\
				1220--1240 & & $0.13686\pm0.00040$\\
				1240--1260 & & $0.13761\pm0.00042$\\
				1260--1280 & & $0.13739\pm0.00044$\\
				1280--1300 & & $0.13714\pm0.00043$\\
				1300--1320 & & $0.13733\pm0.00040$\\
				1320--1340 & & $0.13711\pm0.00038$\\
				1340--1360 & & $0.13736\pm0.00041$\\
				1360--1380 & & $0.13798\pm0.00038$\\
				1380--1400 & & $0.13730\pm0.00038$\\
				1400--1420 & & $0.13827\pm0.00036$\\
				1420--1440 & & $0.13817\pm0.00039$\\
				1440--1460 & & $0.13783\pm0.00039$\\
				1460--1480 & & $0.13744\pm0.00041$\\
				1480--1500 & & $0.13754\pm0.00037$\\
				1500--1520 & & $0.13703\pm0.00050$\\
				1520--1540 & & $0.13697\pm0.00039$\\
				1540--1560 & & $0.13667\pm0.00038$\\
				1560--1580 & & $0.13679\pm0.00040$\\
				1580--1600 & & $0.13710\pm0.00041$\\
				1600--1620 & & $0.13741\pm0.00039$\\
				1620--1640 & & $0.13636\pm0.00043$\\
				1640--1660 & & $0.13650\pm0.00043$\\
							
				\hline
			\end{tabular}
			
			\begin{tablenotes}
			\item {\bf Note.} 
			 \item \textsuperscript{a}Narrow wavelength bins centered on the alkali (Na and K) absorption lines.
			\end{tablenotes}
		\end{center}
	\end{threeparttable}
\end{table}

In our pipeline, the transit shape $f(t)$ is calculated using the BATMAN package \citep{batman}. For the global broadband light-curve analysis, we fit for a separate transit depth ($R_{p}/R_{*}$) in each of the five bandpasses (STIS G430L, STIS G750L, WFC3 G141, IRAC 3.6~$\mu$m, and IRAC 4.5~$\mu$m), along with a single set of transit geometry parameters ($a/R_{*}$, $b$) and transit ephemerides ($T_{0}$, $P$) for all light curves. 

The log-likelihood function for our joint light-curve fits is
\begin{align}\label{logL}\log L = \sum\limits_{V=1}^{N}\Bigg\lbrack -n_{V}&\log {\sqrt{2\pi}}  \sigma_{V}\notag \\& - \frac{1}{2}\sum\limits_{i=1}^{n_V}\frac{[D_{V}(t)-S_{V}(t)\cdot f_{V}(t)]^2}{\sigma_{V}^{2}}\Bigg\rbrack,\end{align}
where the outer summation goes over all $N=8$ visits. For each visit $V$, $n_{V}$ is the number of data points $D_{V}$, $S_{V}$ is the appropriate instrumental systematics model (Section~\ref{systematics}), $f_{V}$ is the transit light-curve model, and $\sigma_{V}$ is a free photometric noise parameter. We have introduced an independent noise parameter for each visit to account for differences in the level of scatter across the various transit light curves. The best-fit values of the noise parameters establish conservative estimates of the photometric uncertainty on each data point.

The ExoTEP pipeline simultaneously computes the best-fit values and $\pm1\sigma$ uncertainties for all astrophysical and systematics model parameters using the affine-invariant Markov Chain Monte Carlo (MCMC) ensemble sampler \texttt{emcee} \citep{emcee}. To facilitate convergence of the chains, we initialize the global fit with the best-fit values calculated by fitting each transit individually. The global transit fit contains a total of 53 free astrophysical, systematics, and noise parameters. We use $53\times4=212$ walkers and chain lengths of 20,000 steps, discarding the first 60\% of each chain when computing the posterior distributions of the fit parameters. To check for convergence, we run the fit five times and ensure that the parameter estimates are consistent across the five runs at better than the $0.1\sigma$ level. The results of our global transit light-curve analysis are listed in Table~\ref{tab:fit}. Plots of the best-fit transit light curves and their corresponding residuals are shown in Figures~\ref{wfcwlc}--\ref{spitzerwlc}. 

\begin{figure}[t!]
\begin{center}
\includegraphics[width=9cm]{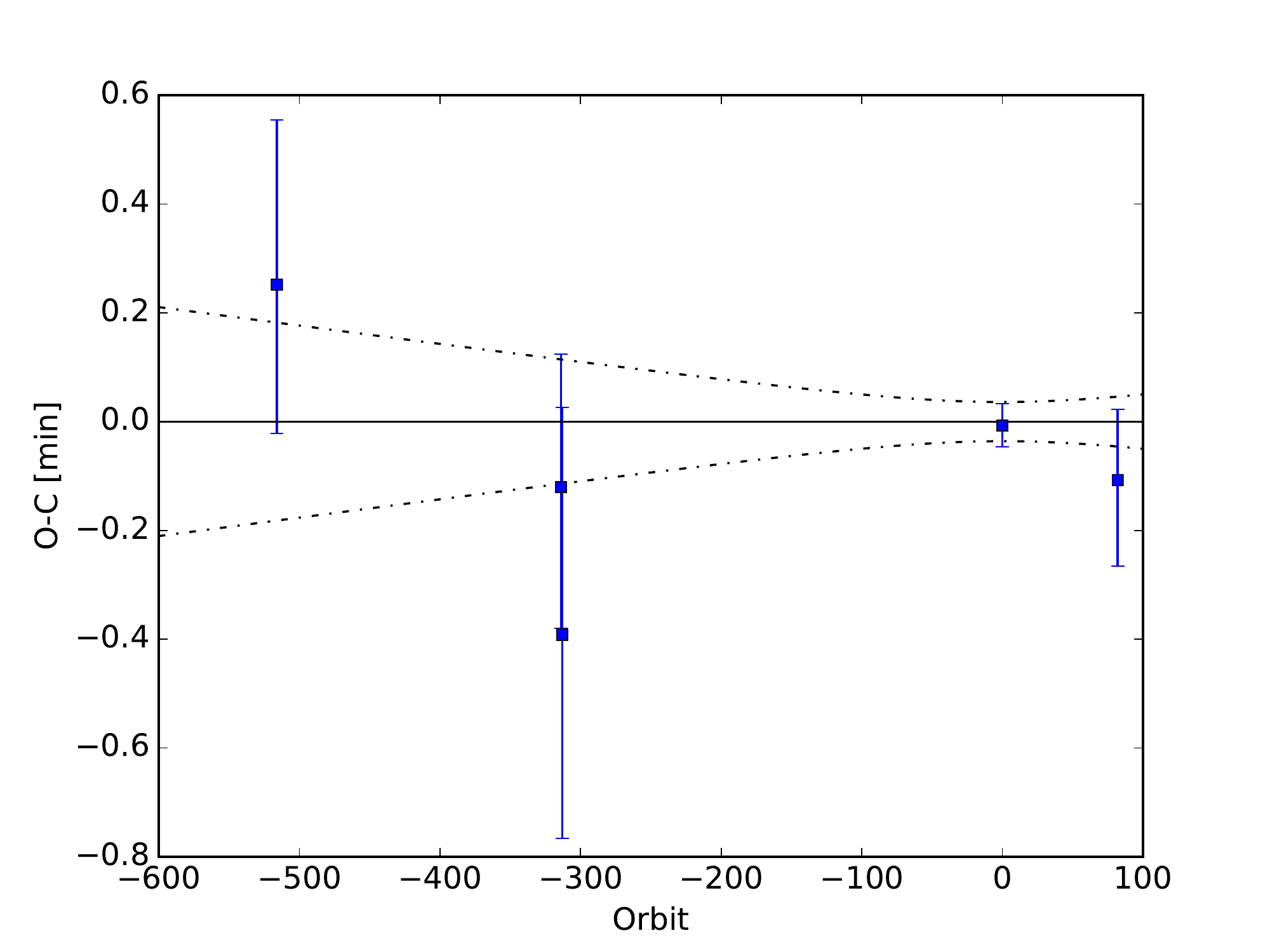}
\end{center}
\caption{Observed minus calculated transit time plot showing the best-fit individual WFC3 G141 and IRAC 3.6 and 4.5~$\mu$m transit times (blue points) relative to the best-fit transit ephemeris derived from the global broadband transit fit (black curves). The STIS transit times are not included because the light curves from those visits do not cover ingress or egress, resulting in significantly larger transit time uncertainties.} \label{transitcompare}
\end{figure}

Our global fits assume a single transit ephemeris across all visits, as well as a common transit depth for visits observed in the same bandpass. To validate this treatment, we also analyze each visit individually in order to compare the best-fit transit timings with the global best-fit transit ephemeris and ensure consistent transmission spectrum shapes among the visits. Figure~\ref{transitcompare} shows the calculated transit times for individual visits relative to the best-fit global transit ephemeris; only visits with full transit coverage or partial coverage including ingress and egress are included. All of the individual transit times agree with the global ephemeris at better than the $1\sigma$ level, ruling out any statistically significant transit timing variation.

\begin{figure}[t!]
\begin{center}
\includegraphics[width=9cm]{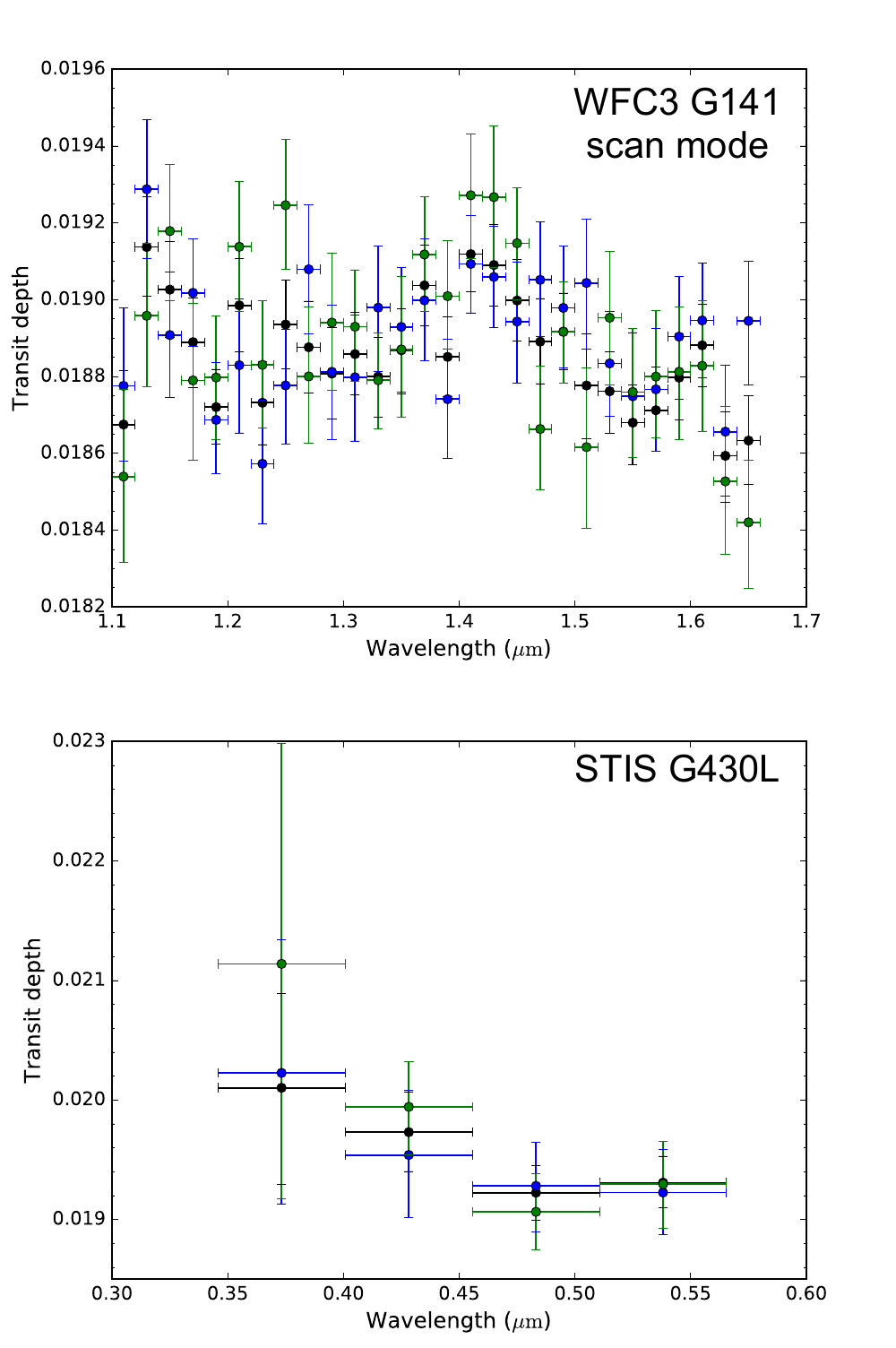}
\end{center}
\caption{Comparison plot showing the transmission spectra derived from the individual visits in the WFC3 G141 (scan mode only) and STIS G430L bandpasses (blue and green points) alongside the corresponding spectra computed from the joint analysis (black points). The individual spectra agree well with the joint spectrum across all wavelengths. } \label{spectrumcompare}
\end{figure}

\subsection{Spectroscopic Light-curve Fits}\label{subsec:speclc}

When fitting the individual spectroscopic light curves in the STIS G430L, STIS G750L, and WFC3 G141 bandpasses, we fix the transit geometry parameters and transit ephemeris to the best-fit values from the global broadband transit analysis (Table~\ref{tab:fit}), with the transit depth being the only free astrophysical parameter. 

The ExoTEP pipeline offers a choice of three methods for defining the instrumental systematics model for the constituent spectroscopic light curves. The first method utilizes the full systematics model for the corresponding instrument, computing the best-fit instrumental systematics parameters for each spectroscopic light curve independently from the broadband light curve. The other methods apply a common-mode correction to the spectrophotometric series prior to fitting, dividing each series by either (1) the best-fit broadband systematics model \citep[e.g.,][]{kreidberg}, or (2) the ratio of the uncorrected broadband photometric series and the best-fit broadband transit model \citep[e.g.,][]{deming2013}. 

Performing a precorrection on the spectroscopic light curves takes advantage of the more well-defined systematics model derived using the high signal-to-noise broadband light curves. This technique also enables us to use fewer systematics parameters in the individual spectroscopic light-curve fits, which typically results in tighter constraints on the best-fit transit depths. We account for residual systematic flux variations in the spectroscopic light curves using a simplified model,
\begin{equation}\label{specsystematics}S_{\mathrm{spec}}(t) = c+v\cdot(x-x_{0}),\end{equation}
which describes a linear function with respect to the measured subpixel shifts $x-x_{0}$ in the dispersion direction relative to the first exposure in the time series, with $c$ and $v$ being the offset and slope parameters, respectively.

To demonstrate consistency in the transmission spectrum shape between separate observations in the same bandpass, we first analyze the spectroscopic light curves of individual visits. Figure~\ref{spectrumcompare} shows the transmission spectra of the individual WFC3 G141 scan mode and STIS G430L visits plotted with the corresponding spectra derived from the joint analysis. In both cases, there is good agreement between the individual transit depths in each wavelength bin, and the spectrum shapes are consistent across the visits. In particular, each WFC3 G141 scan mode visit spectrum shows a discernible absorption feature at 1.4~$\mu$m. It is also important to note that this feature was not detected in the older stare mode data analyzed in \citet{line}, which underscores the significant improvement in sensitivity provided by the scan mode observations.

Using the same log-likelihood expression as in our global broadband transit light-curve fit (Eq.~\eqref{logL}), we then fit all visits in a given bandpass jointly, letting the systematics model and photometric noise parameters vary independently for each light curve. For the STIS G430L and G750L spectroscopic light curves, in line with similar previous studies \citep[e.g.,][]{singstis}, we find that the shapes of the systematic trends vary significantly across the various wavelength bins, necessitating the use of the full systematics model. Meanwhile, the HST WFC3 systematics  are largely independent of wavelength and detector position, and we find that the two precorrection strategies described above result in fits of comparable quality. In this paper, we report the best-fit depths derived from using the latter of the two precorrection methods (i.e., dividing the ratio of the uncorrected flux and the best-fit transit model from the broadband light curve).

\begin{figure*}[t!]
\begin{center}
\includegraphics[width=\linewidth]{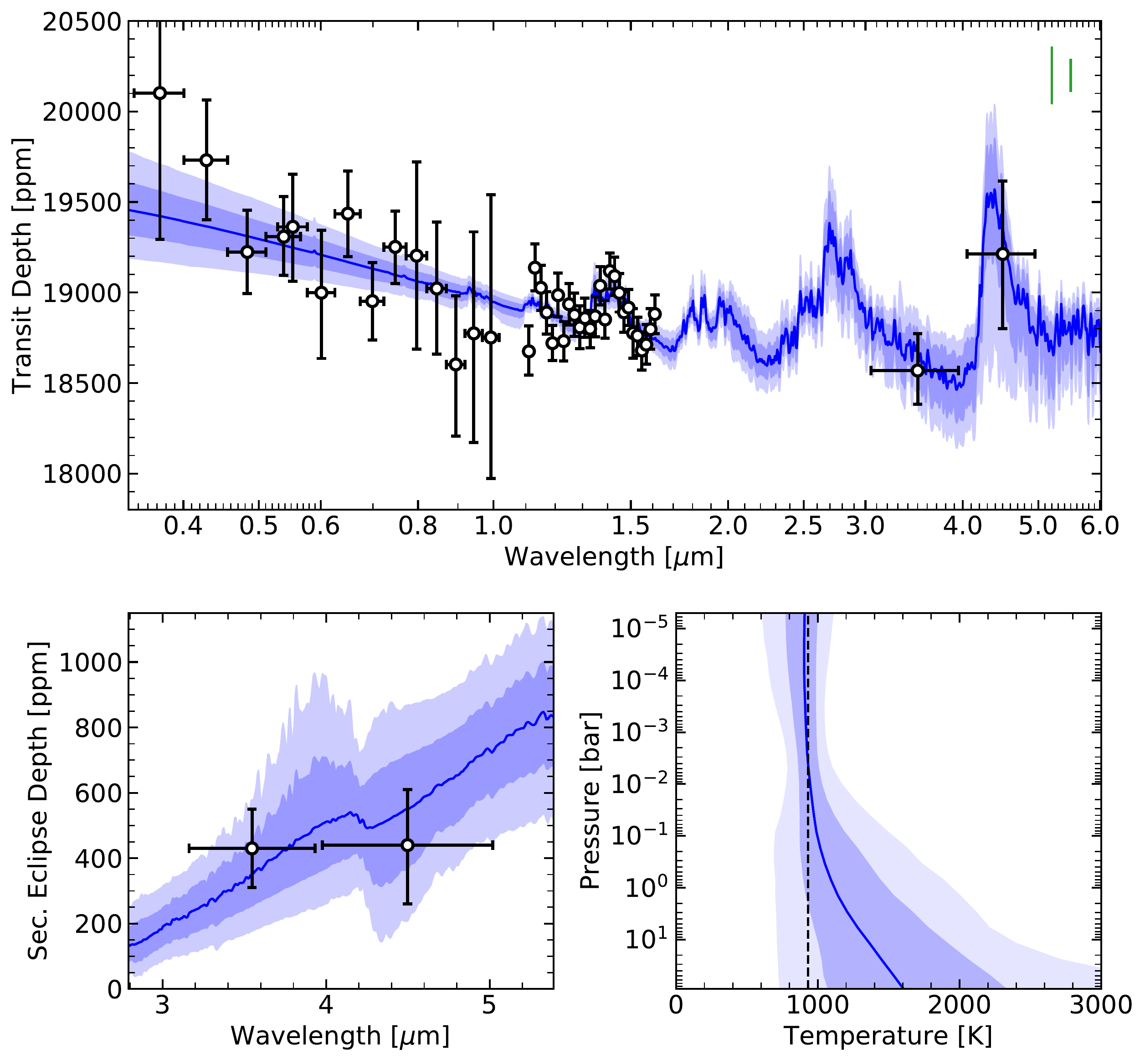}
\end{center}
\caption{Top: transmission spectrum of HAT-P-12b computed from our global broadband and spectroscopic transit light-curve analysis (black circles). Model transmission spectra from our atmospheric retrievals are also plotted for comparison. The shaded regions indicate 1$\sigma$ and 2$\sigma$ credible intervals in the retrieved spectrum (medium and light blue, respectively) relative to the median fit (dark blue line). The main features of the transmission spectrum are the Rayleigh scattering slope at visible wavelengths and a weak water vapor feature at 1.4~$\mu$m; both of these features are well modeled by the retrieval. The vertical green bars in the top right corner indicate the variation in transit depth corresponding to one atmospheric scale height in the best-fit model (184~ppm) and a solar composition atmosphere (320~ppm). Bottom left: same as top panel but for the emission spectrum derived from the Spitzer IRAC secondary eclipses. The relatively low-precision broadband secondary eclipse depths are consistent with a wide range of emission spectrum shapes. Bottom right: median temperature--pressure profile from the retrieval (solid blue curve), along with $1\sigma$ and $2\sigma$ bounds. The vertical dashed line indicates the equilibrium temperature for complete heat redistribution assuming a planetary Bond albedo of $A=0.1$.} \label{spectrum}
\end{figure*}

The results of our spectroscopic light-curve fits are listed in Table~\ref{tab:fit2}. The best-fit transit light curves and associated residuals are plotted in the Appendix for each of the HST STIS and WFC3 visits. When experimenting with different wavelength bin widths (10--40~nm), we get consistent transmission spectrum shapes. Visual inspection of the systematics-corrected light curves does not reveal any salient outliers or residual uncorrected systematics trends. We combine the transit depths from the spectroscopic light-curve fits with the broadband Spitzer IRAC transit depths to construct the full transmission spectrum of HAT-P-12b, which is plotted in Figure~\ref{spectrum}. The transit depths for the narrow wavelength bins in the main alkali absorption regions are consistent with the depths measured in the wider bins spanning those regions, indicating a nondetection of the alkali absorption; these data points are not shown in the transmission spectrum plot. The primary features of the transmission spectrum are the Rayleigh slope extending through the optical bandpasses and a small absorption feature around 1.4~$\mu$m indicative of water vapor.  These observations together suggest the presence of both uniform clouds and fine-particle scattering in the atmosphere of HAT-P-12b.

The shape of the transmission spectrum at visible wavelengths matches the results of previous analyses of the HST STIS data by \citet{singstis} and \citet{alexoudi}. It is worth mentioning that an earlier study of HAT-P-12b's atmosphere using ground-based broadband photometry produced a flat transmission spectrum throughout the visible wavelength range \citep{mallonn}, consistent with an opaque layer of clouds as opposed to Rayleigh scattering. This discrepancy was discussed in \citet{alexoudi} and attributed to uncertainties in the inclination and semi-major axis of HAT-P-12b's orbit, which are correlated with transit depths and can yield wavelength-dependent shifts that alter the apparent transmission spectrum slope in the optical. 

When assuming different values of $i$ and $a/R_{*}$ in reanalyzing the \citet{mallonn} light curves, \citet{alexoudi} were able to recover a discernible Rayleigh scattering slope in the visible transmission spectrum. In our global fit, we take advantage of the well-sampled ingress and egress from the scan mode HST WFC3 and Spitzer light curves to place much narrower constraints on $i$ and $a/R_{*}$ than these earlier studies. Therefore, our results are a robust validation of the previously published reports of a negative slope in the visible transmission spectrum of HAT-P-12b.

\subsection{Secondary Eclipses}\label{subsec:ecl}

The eclipse light curve is defined in the same way as a transit light curve but without the limb-darkening effect. We utilize the same modified PLD instrumental systematics model to account for the Spitzer IRAC intrapixel sensitivity variations (Eq.~\eqref{pld}). For each eclipse observation, we select the optimal aperture, photometric parameters, binning, and trimming by fitting the eclipse light curve individually, fixing the transit geometry parameters ($a/R_{*}$, $b$) and transit ephemerides ($T_{0}$, $P$) to the best-fit values from the global broadband transit light-curve analysis (Table~\ref{tab:fit}).

When performing individual eclipse fits on the relatively low signal-to-noise data, we facilitate comparison between different versions of the photometry/binning/trimming by fixing the time of eclipse to an orbital phase of 0.5. The orbital phase here is defined relative to the best-fit ephemeris from the global transit fit. To correct for any residual flux ramps at the start of the data, we also experiment with including an exponential factor $(1 - a_{1}e^{-t_{i}/a_{2}})$ in the systematics model, where $a_{1}$ and $a_{2}$ are the amplitude and time constant, respectively, and $t_{i}$ is the time elapsed since the beginning of the time series. Following \citet{wong2,wong3}, we choose the photometric series that produce the lowest residual scatter. Only for the first 3.6~$\mu$m eclipse dataset does the inclusion of a ramp appreciably improve the fit (i.e., minimizes the value of the Bayesian Information Criterion). 

\begin{figure}[t!]
\begin{center}
\includegraphics[width=9cm]{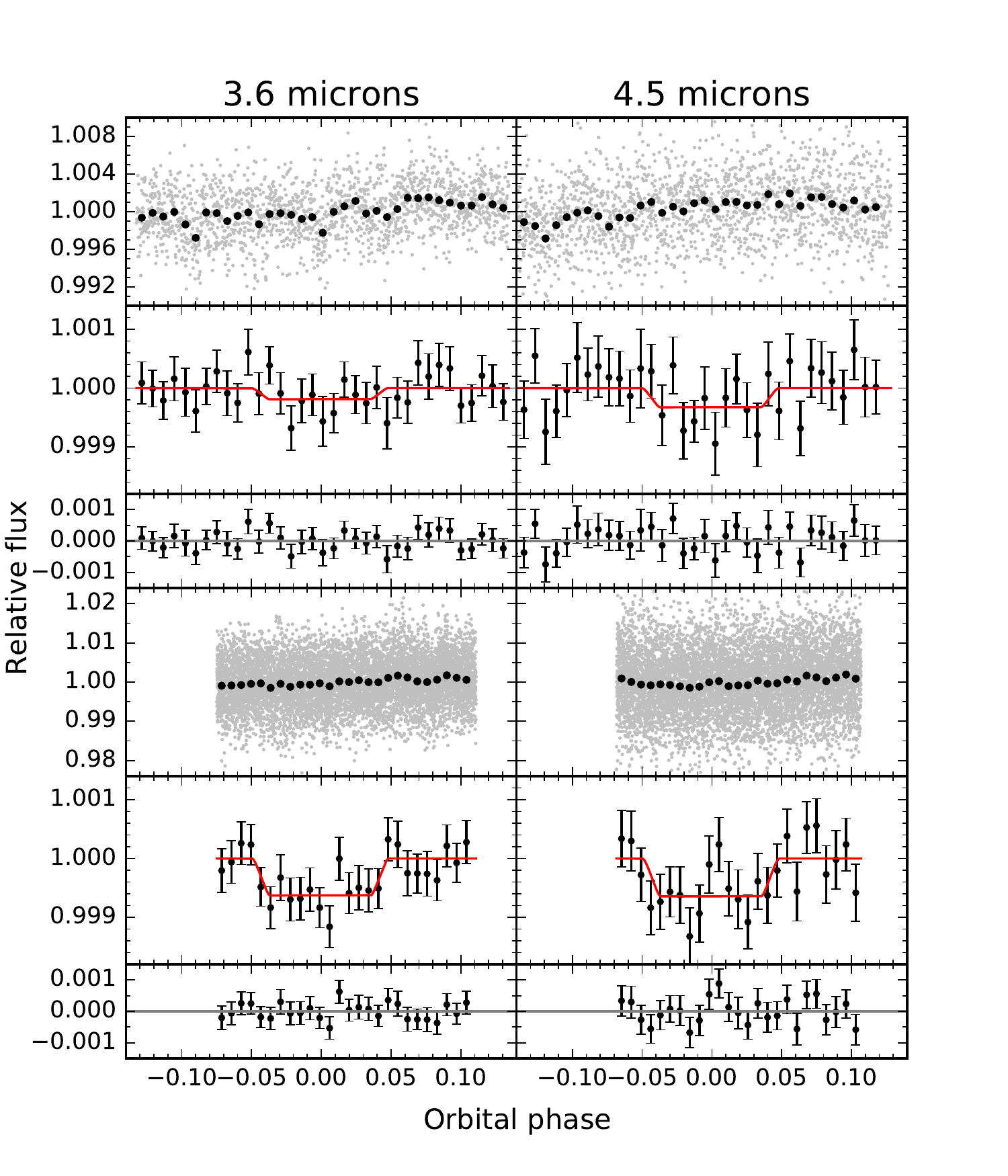}
\end{center}
\caption{Left: plots of the two Spitzer 3.6~$\mu$m secondary eclipses, binned in 5 minute intervals. In the top panels, the unbinned photometric series is shown in gray, with the binned data overplotted in black. The middle panels show the corrected light curve with the intrapixel sensitivity effect removed. The best-fit eclipse light curve is overplotted in red. The corresponding residuals from the fit are shown in the bottom panels. The error bars shown are the standard deviation of the residuals from the best-fit light curve, scaled by the square root of the number of points in each 5 minute bin. Right: analogous plots for the two Spitzer 4.5~$\mu$m secondary eclipses.} \label{eclipses}
\end{figure}

We also carry out a global analysis of all four secondary eclipse observations. In this fit, we allow the instrumental systematics parameters for each dataset to vary independently while assuming common 3.6 and 4.5~$\mu$m eclipse depths and center of eclipse phase as free parameters. The results of our individual and global eclipse fits are listed in Table~\ref{tab:eclipse}. The raw and systematics-corrected eclipse light curves are shown in Figure~\ref{eclipses}. From the individual fits, we only find marginal eclipse detections for the full array 3.6 and 4.5~$\mu$m visits ($<1.5\sigma$), while the more recent subarray observations yield more robust detections ($>2.5\sigma$). The best-fit eclipse phase from the combined analysis is consistent with a circular orbit, and the global 3.6 and 4.5~$\mu$m depths are statistically consistent with each of the individual best-fit eclipse depths at better than the $1.1\sigma$ level.

\begin{table}[t!]
\small
	\centering
	\begin{threeparttable}
		\caption{Secondary Eclipse Fit Results} \label{tab:eclipse}	
		\renewcommand{\arraystretch}{1.2}
		\begin{center}
			\begin{tabular}{ l m{0.1cm} c m{0.1cm} c} 
				\hline\hline
				Eclipse  & & Depth (\%) & &  Phase \\
				 \hline
				3.6~$\mu$m & & &  & \\
				\quad Eclipse 1 & & $0.019\pm0.017$ & &$\equiv0.5$\textsuperscript{a} \\
				\quad Eclipse 2 & &$0.064^{+0.017}_{-0.018}$&& $\equiv0.5$\\
			  	\quad Global\textsuperscript{b} & &$0.042\pm0.013$ & &$0.5009^{+0.0026}_{-0.0014}$ \\
				4.5~$\mu$m && & &\\
				\quad Eclipse 1 && $0.032\pm0.024$ && $\equiv0.5$ \\
				\quad Eclipse 2 && $0.066^{+0.027}_{-0.026}$ && $\equiv0.5$ \\
				 \quad Global\textsuperscript{b} && $0.045^{+0.017}_{-0.019}$ & &$0.5009^{+0.0026}_{-0.0014}$ \\
				\hline
			\end{tabular}
			
			\begin{tablenotes}
			 \item {\bf Notes.} 
			 \item \textsuperscript{a}The eclipse phase was fixed at 0.5 for all individual eclipse fits, assuming the best-fit orbital ephemeris from the global transit fit (Table~\ref{tab:fit}).
			 \item \textsuperscript{b}Computed from a simultaneous fit of all four eclipses.
			\end{tablenotes}
		\end{center}
	\end{threeparttable}
\end{table}

\section{Atmospheric Retrieval}\label{sec:disc}


We simultaneously interpret the full transmission and emission spectra presented in this work to deliver quantitative constraints on the atmosphere of HAT-P-12b using the SCARLET atmospheric retrieval framework  \citep[][]{bennekeseager1,bennekeseager2,kreidberg,knutson2014,benneke2015,benneke2019}. Employing SCARLET's chemically consistent mode, we define the atmospheric metallicity, the C/O ratio, the cloud properties, and the vertical temperature structure as free parameters. SCARLET then determines their posterior constraints by combining a chemically-consistent atmospheric forward model with a Bayesian MCMC analysis. We perform the retrieval analysis with 100 walkers using uniform priors on all the parameters and run the chains well beyond formal convergence to obtain smooth posterior distribution even near the $3\sigma$ contours.


To evaluate the likelihood for a particular set of atmospheric parameters, the SCARLET forward model in chemically consistent mode first computes the molecular abundances in chemical and hydrostatic equilibrium and the opacities of molecules and Mie-scattering clouds \citep[][]{bennekeseager2}. The elemental composition in the atmosphere is parameterized using the atmospheric metallicity, [M/H], and the atmospheric C/O ratio. We employ log-uniform priors, and we consider the line opacities of H$_2$O, CO, and CO$_2$ from HiTemp \citep{rothman} and CH$_4$, NH$_3$, HCN, H$_2$S, C$_2$H$_2$, O$_2$, OH, PH$_3$, Na, K, TiO, SiO, VO, and FeH from ExoMol \citep{tennyson}, as well as the collision-induced absorption of H$_2$ and He.


Following \citet[][]{benneke2019}, we use a three-parameter Mie-scattering cloud description for the retrieval analysis defining the mean particle size $R_{\mathrm{part}}$, the pressure level $P_{\tau=1}$ at which the clouds become optically opaque to grazing starlight at 1.5~$\mu$m, and the scale height of the cloud profile relative to the gas pressure scale height $H_{\mathrm{part}}/H_{\mathrm{gas}}$ as free parameters. All free parameters are allowed to vary independently in the retrieval. When calculating the cloud opacity, the retrieval is agnostic to the particular composition of the spherical cloud particles, considering only their size and vertical distribution; the former is assumed to be a logarithmic Gaussian distribution with a fixed width of $\sigma_{R}=1.5$. This three-parameter cloud description is motivated by the information content of transmission spectra and captures the wavelength-dependent opacities of a wide range of finite-sized cloud particles near the cloud deck in a highly orthogonal way, ideal for retrieval \citep[][]{benneke2019}. It reduces to Rayleigh hazes in the limit of small particles and a gray cloud deck for large particles while simultaneously allowing for any finite-sized Mie-scattering particles in between. We employ log-uniform priors on the three cloud parameters.
 

Our temperature structure is parameterized using the five-parameter analytic model from \citet{parmentier2014} augmented with a constraint on the plausibility of the total outgoing flux. Given the relatively weak constraints on the atmospheric composition, we conservatively ensure the plausibility of the temperature structure by enforcing that the wavelength-integrated outgoing thermal flux is consistent with the stellar irradiation, a Bond albedo between 0 and 0.7, and heat redistribution values between full heat redistribution across the planet and no heat redistribution. In the retrieval, we parameterize only one temperature structure for both the dayside and the terminator because the retrieved temperature uncertainties are hundreds of K and the precision of the transmission spectrum does not justify additional parameters describing the terminator temperature structure separately.


Finally, high-resolution synthetic transmission and emission spectra are computed using line-by-line radiative transfer and integrated over the appropriate instrument response functions before being compared to the observations. Sufficient wavelength resolution in the synthetic spectra is ensured by repeatedly verifying that the likelihood for a given model is not significantly affected by the finite wavelength resolution ($\Delta\chi^2<0.001$). Reference models are computed at $\frac{\lambda}{\Delta\lambda}=250,000$.

\begin{table}[t!]
	\centering
	\begin{threeparttable}
		\caption{HAT-P-12b Atmospheric Retrieval Results} \label{retrievalresults}	
		\renewcommand{\arraystretch}{1.2}
		\begin{center}
			\begin{tabular}{ l  c  c  c  } 
				\hline\hline
				Parameter & Value & Unit \\
				\hline
				Atmospheric metallicity, $\log M$ & $2.43^{+0.33}_{-0.60}$ & x solar \\
				Atmospheric C/O ratio & $0.48_{-0.37}^{+0.10}$ & $\dots$ \\
				Mean particle size,\textsuperscript{a} $\log R_{\mathrm{part}}$ & $-1.47^{+0.45}_{-0.36}$ & $\mu$m \\
				Opacity pressure level,\textsuperscript{b} $\log P_{\tau=1}$ & $0.38^{+2.23}_{-1.18}$ & mbar \\
				Relative cloud scale height,\textsuperscript{c}  & &\\
				\qquad$\log H_{\mathrm{part}}/H_{\mathrm{gas}}$ & $0.12^{+0.30}_{-0.34}$ & $\dots$ \\
										
				\hline				
			\end{tabular}
			
			\begin{tablenotes}
				\small
				\item {\bf Notes.}
				\item \textsuperscript{a}Mean particle size, assuming a logarithmic Gaussian distribution with fixed width of $\sigma_{R}=1.5$
				\item \textsuperscript{b}Pressure at transmission optical depth of unity at 1.5~$\mu$m
				\item \textsuperscript{c}Scale height of cloud profile relative to gas pressure scale height
				
			\end{tablenotes}
		\end{center}
	\end{threeparttable}
\end{table}


\begin{figure*}[t!]
\begin{center}
\includegraphics[width=\linewidth]{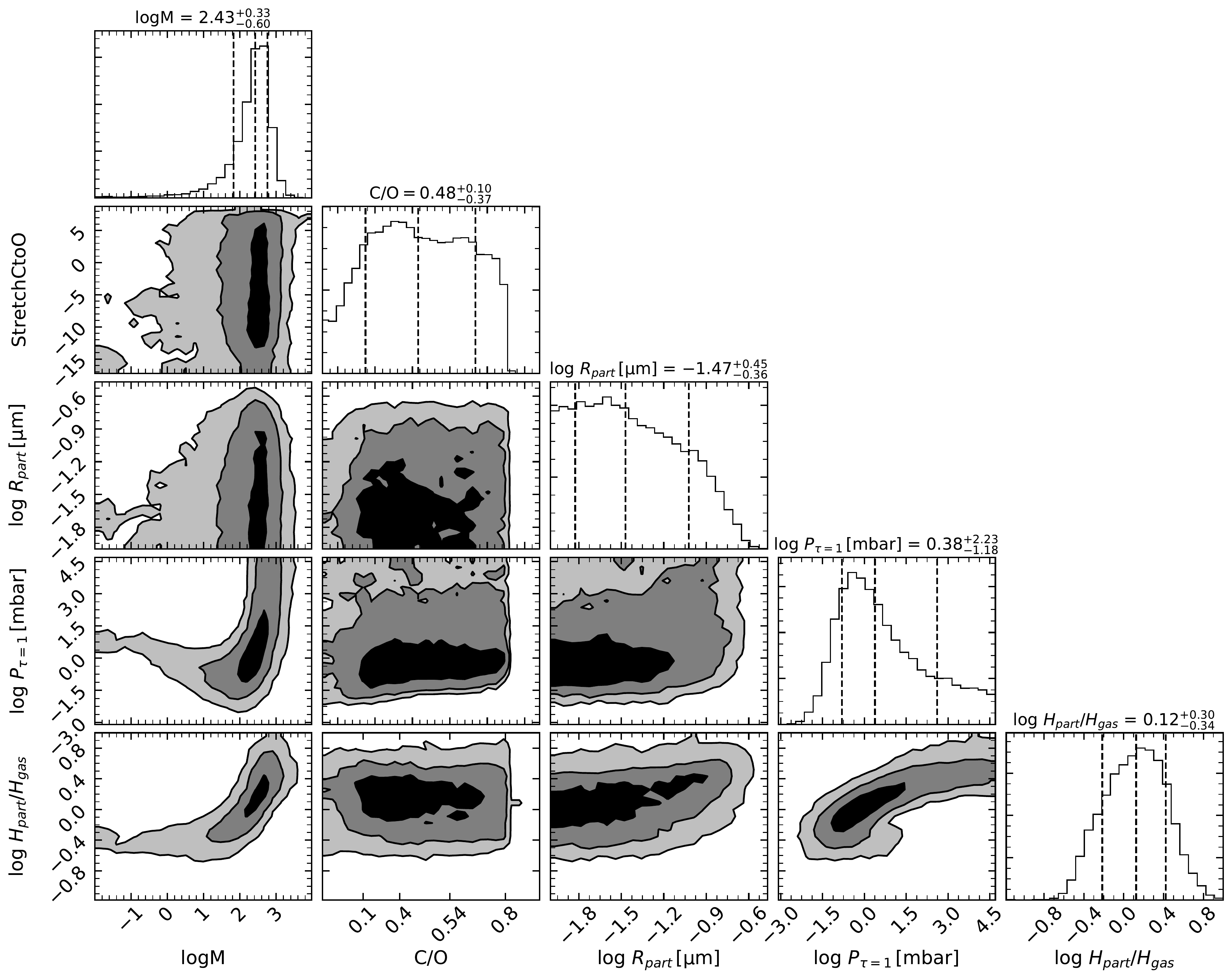}
\end{center}
\caption{Triangle plot showing the one- and two-parameter marginalized posteriors from the SCARLET atmospheric retrieval of the HAT-P-12b transmission and emission spectra. The black, dark gray, and light gray regions denote $1\sigma$, $2\sigma$, and $3\sigma$ regions.} \label{triangle}
\end{figure*}

\begin{figure}[h!]
\begin{center}
\includegraphics[width=\linewidth]{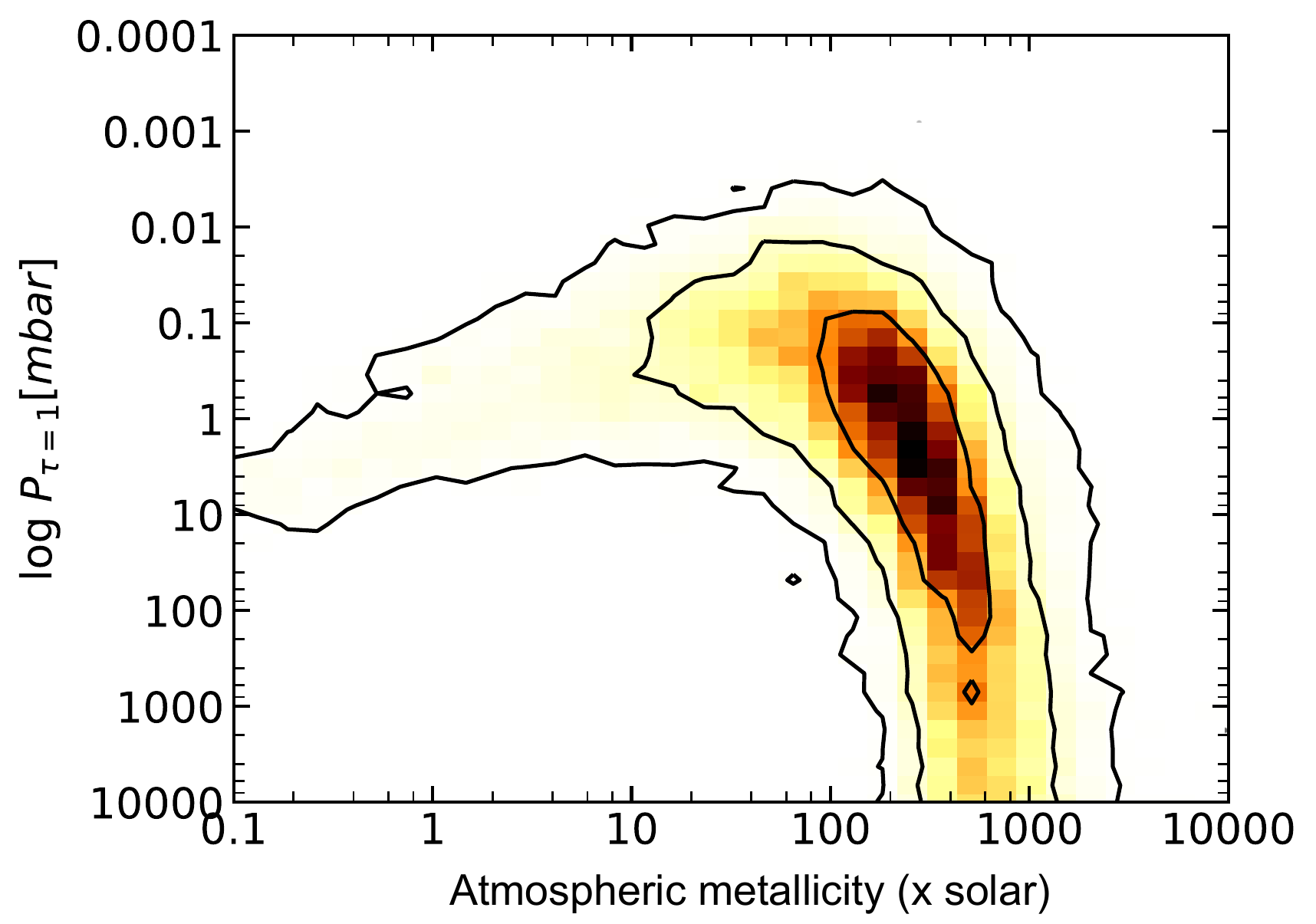}
\end{center}
\caption{The 2D posterior of cloud-top pressure $\log(P_{\tau=1})$ vs. atmospheric metallicity from the atmospheric retrieval. The solid black lines indicate $1\sigma$, $2\sigma$, and $3\sigma$ bounds. The HAT-P-12b transmission spectrum is consistent with both cloudy atmospheres spanning a broad range of metallicities and clear atmospheres with highly enhanced metallicities.} \label{cloudmetal}
\end{figure}

\subsection{Retrieval results}

We run a set of retrievals that assume chemical and thermal equilibrium, setting the atmospheric metallicity $\log M$, atmospheric C/O ratio, and cloud properties ($R_{\mathrm{part}}$, $P_{\tau=1}$, and $H_{\mathrm{part}}/H_{\mathrm{gas}}$) as free parameters. The range of representative atmospheric models derived from the retrievals is illustrated in Figure~\ref{spectrum}. The median atmospheric model is shown by the blue curve, and the $1\sigma$ and $2\sigma$ credible intervals are indicated by the shaded regions. In short, both the  observed transmission and emission spectra are well fit across all wavelengths by cloudy atmospheres with cloud-top pressures between 0.2~mbar and 0.4~bar ($1\sigma$ bounds) and supersolar metallicities. The data favor submicron cloud particle sizes, and the posterior spans most of the assumed prior range below $\sim$200~nm. Crucially, the retrieved particle sizes cover the range necessary to produce the Rayleigh scattering in the optical evident in the transmission spectrum, consistent with a previous retrieval of the HAT-P-12b atmosphere \citep{barstow}. The list of parameter estimates is given in Table~\ref{retrievalresults}.

The full triangle plot displaying all one- and two-parameter marginalized posteriors is shown in Figure~\ref{triangle}. Of particular interest is the degeneracy between cloud-top pressure and atmospheric metallicity, which is shown separately in Figure~\ref{cloudmetal}. Overall, the atmospheric metallicity is not well constrained: the L-shaped posterior indicates that while the data are largely consistent with cloudy atmospheres spanning a wide range of supersolar metallicities, clear atmospheres with strongly enhanced metallicities above 100 times solar cannot be ruled out at the $1\sigma$ level. This degeneracy is a common feature in atmospheric retrievals of exoplanet transmission spectra with weak or undetected 1.4~$\mu$m water features \citep[e.g., HAT-P-11b;][]{fraine2014}, where the small magnitude of the water absorption can be caused either by attenuation due to the presence of clouds or by an intrinsically weak absorption from a hydrogen-depleted atmosphere with high mean molecular weight.

Core accretion models predict a trend of increasing bulk metallicity with decreasing planet mass \citep[e.g.,][]{mordasini2012,fortney2013}, and most known gas giant exoplanets have supersolar bulk metallicities \citep[e.g.,][]{thorngren2016}. Meanwhile, the relationship between bulk and atmospheric metallicity is more complex. From planet evolution and interior structure modeling, \citet{thorngren2019} predicted a 95\% atmospheric metallicity upper limit of 82.3 for HAT-P-12b, broadly consistent with the results of our atmospheric retrievals and the corresponding bulk metallicity of the planet. Further enrichment of the atmospheric metallicity can result from secondary processes such as core erosion \citep[e.g.,][]{wilson2012,madhusudhan2016} and accretion of solid material during the late stages of planet formation \citep[e.g.,][]{pollack1996}. The atmospheric metallicity of HAT-P-12b is also comparable to similarly sized sub-Saturn planets, such as WASP-39b (100--200$\times$ solar; \citealt{wakeford2017}) and WASP-127b (10--40$\times$ solar; \citealt{spake2019}). Given this context, the elevated metallicity of HAT-P-12b is not entirely unexpected.

Another notable result from the retrievals is the near-solar atmospheric C/O ratio of $0.48_{-0.37}^{+0.10}$, with a $3\sigma$ upper limit at roughly 0.83. The presence of a water vapor absorption feature at 1.4~$\mu$m rules out carbon-dominated atmospheres, because the formation of H$_{2}$O becomes disfavored as C/O approaches unity. The absence of a 1.15~$\mu$m absorption in the WFC3 bandpass comparable in magnitude to the observed 1.4~$\mu$m feature also supports the conclusion of an oxygen-dominated chemistry by eliminating CH$_{4}$ as the molecular species responsible for the near-infrared absorption features \citep[e.g.,][]{benneke2015}. Methane has a strong absorption feature at around 3.3~$\mu$m, so it follows that the relatively low transit depth measured in the Spitzer 3.6~$\mu$m bandpass in comparison with the 4.5~$\mu$m depth likewise points toward a near-solar C/O ratio.

In addition to the chemical and thermal equilibrium retrievals, we run a set of ``free" atmospheric retrievals that do not assume chemical or thermal equilibrium; instead, the abundance of each molecular gas species is independently varied, in addition to the previously defined parameters describing the clouds. In these runs, we focus on H$_{2}$O, CH$_{4}$, CO, and CO$_{2}$ as the primary atmospheric components to be constrained. We do not find any notable constraints on the abundances of the carbon-bearing species relative to H$_{2}$O.

\begin{figure*}[t]
\begin{center}
\includegraphics[width=\linewidth]{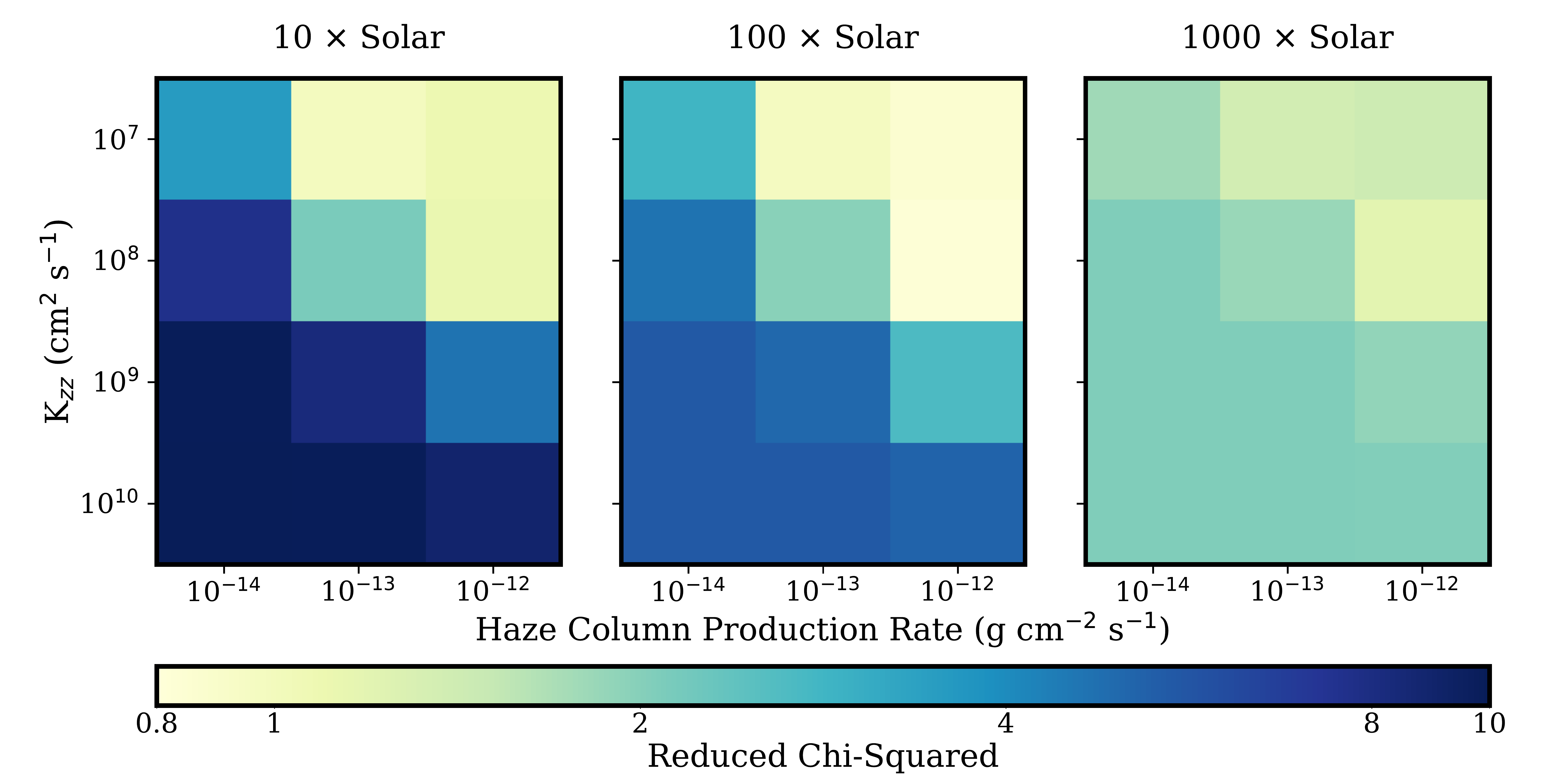}
\includegraphics[width=\linewidth]{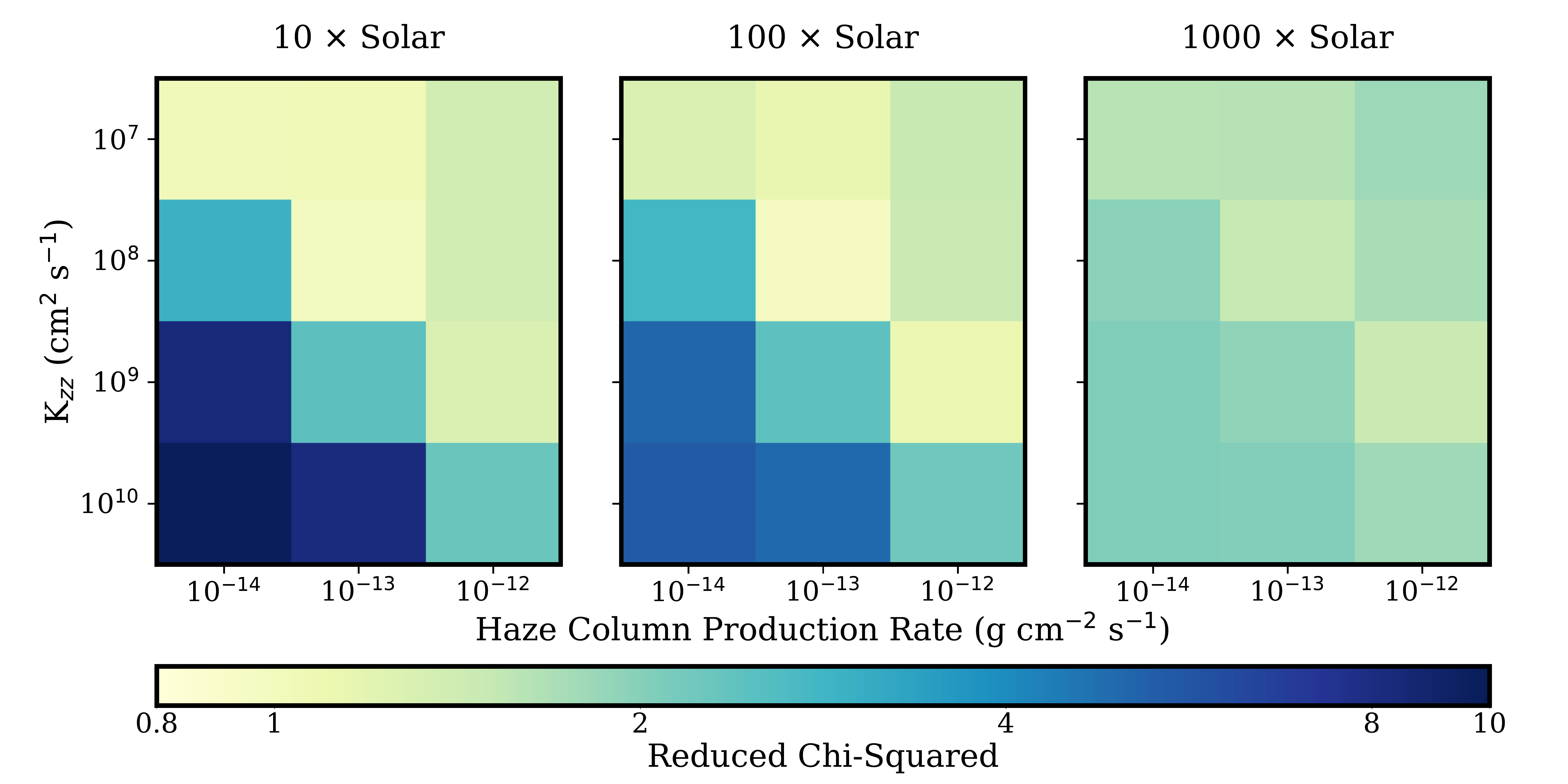}
\end{center}
\caption{Top: grid of RCS values for all 36 CARMA runs that included photochemical haze particles composed of tholins. Bottom: same as top panel but for soot model runs. The best-performing model runs for tholins and soot assume an atmospheric metallicity of 100$\times$ solar and an eddy diffusion coefficient of $K_{zz}= 10^{9}$~cm$^2$~s$^{-1}$. For tholins, the model with a haze production rate of $10^{-12}$~g~cm$^{-2}$~s$^{-1}$ best matches the observations, while in the case of soot, a lower rate of $10^{-13}$~g~cm$^{-2}$~s$^{-1}$ is preferred.} \label{RCS}
\end{figure*}

\section{Comparison to Microphysical Cloud Models}\label{sec:carma}

In addition to the atmospheric retrievals presented in the previous section, we use the Community Aerosol and Radiation Model for Atmospheres (CARMA) to simulate condensation clouds and photochemical hazes in the atmosphere of HAT-P-12b. CARMA is a time-stepping cloud microphysics model that computes the bin-resolved particle size distributions of aerosols as a function of altitude in planetary atmospheres. CARMA treats aerosol formation and evolution as a kinetic processes, with convergence dictated by balancing the rates of particle nucleation, condensational growth and evaporation, coagulation, and transport via sedimentation, advection, and diffusion calculated from classical theories of cloud physics \citep{pruppacher}. It is thus significantly different from phase equilibrium models, such as \citet{ackermanmarley2001}, which do not consider the time evolution of the rates of microphysical processes. The specific physical formalism used in the model is described in full in the Appendix of \citet{gao2018}.

\begin{figure*}[t!]
\begin{center}
\includegraphics[width=\linewidth]{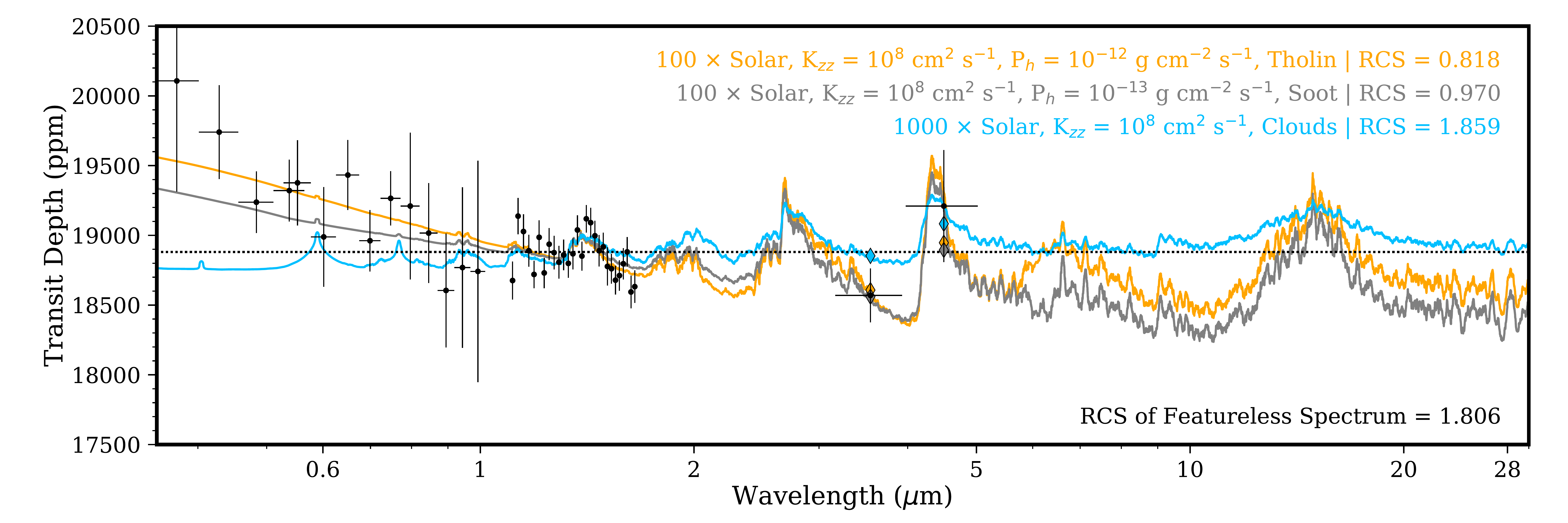}
\end{center}
\caption{Model transmission spectra derived for the best-fitting CARMA condensate cloud, soot, and tholin models. The black points show the observed transmission spectrum. The model parameters for each run are listed in the legend, along with the corresponding RCS value. For comparison, a flat featureless spectrum is also included. While all models match the muted water feature at 1.4~$\mu$m, only the photochemical haze models reproduce the observed Rayleigh scattering slope at optical wavelengths.} \label{carmamodels}
\end{figure*}

By comparing the CARMA simulation results to the observations, we hope to gain a more physical understanding of the processes controlling aerosol distributions.  
In our modeling of the HAT-P-12b atmosphere, we consider both condensate clouds and photochemical hazes separately. We refer the reader to the Appendix for a detailed description of the condensate and aerosol modeling setup in CARMA. For each model run, the temperature--pressure profile of the background atmosphere is set to the best-fit profile from the atmospheric retrieval (Section~\ref{sec:disc} and Figure~\ref{spectrum}). Vertical mixing of condensate or haze particles is driven by eddy diffusion, and we consider eddy diffusion coefficient $K_{zz}$ values of $10^{7}$,  $10^{8}$,  $10^{9}$, and  $10^{10}$~cm$^2$~s$^{-1}$.  The atmospheric metallicity is set to 10$\times$, 100$\times$, or 1000$\times$ solar; adjusting the metallicity affects the initial abundance of condensate species in the model, as well as the atmospheric scale height. 

Given the uncertainties in the specific chemical pathways and efficiencies of haze production, CARMA does not carry out an \textit{ab initio} haze formation calculation but instead sets the haze production rate as a free parameter. We consider haze production rates of $10^{-14}$, $10^{-13}$, and $10^{-12}$~g~cm$^{-2}$~s$^{-1}$ at a pressure of 1~$\mu$bar, consistent with the values computed in exoplanet photochemical studies \citep[e.g.,][]{venot2015,lavvas2017,kawashima2018,lines1,adams2019}. We investigate the impact of different haze compositions on the atmospheric opacity by considering different  refractive indices. In particular, we consider both soots, which are expected to survive at the high temperatures of exoplanet atmospheres due to their relatively low volatility, and tholins, which we use as a proxy for lower-temperature organic hazes \citep{morley2015}.

We find that haze models match the observed transmission spectrum much better than condensate cloud models. While many of the condensate cloud models are able to reproduce the shape of the muted water vapor absorption feature at 1.4~$\mu$m, none of them generate the observed steep slope throughout the optical, resulting in reduced $\chi^{2}$ (RCS) values significantly higher than unity. When examining the average particle sizes predicted by the condensate cloud model runs, we find relatively large condensate particles on the order of or exceeding 1~$\mu$m --- too large to allow for Rayleigh scattering in the optical. Meanwhile, the haze models readily reproduce the observed Rayleigh scattering slope. In addition, cloud models that can match the amplitude of the 1.4~$\mu$m water feature are too flat to explain the large offset between the two Spitzer points due to the extensive cloud opacity at 3--5~$\mu$m, while the haze opacity falls off with increasing wavelength sufficiently quickly to allow for larger-amplitude molecular features there. 

Figure~\ref{RCS} shows the RCS values for the full grid of tholin and soot haze models. In both cases, the best-performing run (lowest RCS) has an atmospheric metallicity of 100$\times$ solar and a moderate rate of vertical mixing ($K_{zz}= 10^{8}$~cm$^{2}$~s$^{-1}$). For tholins, the observations are best matched when assuming a haze production rate of $10^{-12}$~g~cm$^{-2}$~s$^{-1}$, whereas for soot, a lower production rate of $10^{-13}$~g~cm$^{-2}$~s$^{-1}$ is preferred, since soots are more absorbing than hazes at the wavelengths of interest \citep{adams2019}. The model transmission spectra derived from the best-fitting condensate cloud, tholin, and soot models are shown in Figure~\ref{carmamodels}. Both of the photochemical haze models match the full set of observations and have RCS values below 1. Meanwhile, the lowest-RCS condensate cloud model performs more poorly than even a featureless flat spectrum. When comparing the soot and tholin spectra, the only salient distinguishing feature is at $\sim$6.5~$\mu$m, where the tholin spectrum displays an additional absorption possibly attributable to double-bonded carbon atoms, double-bonded carbon and nitrogen atoms, and single-bonded amine groups \citep{imanaka2004,gautier2012}.

\begin{figure}[t!]
\begin{center}
\includegraphics[width=\linewidth]{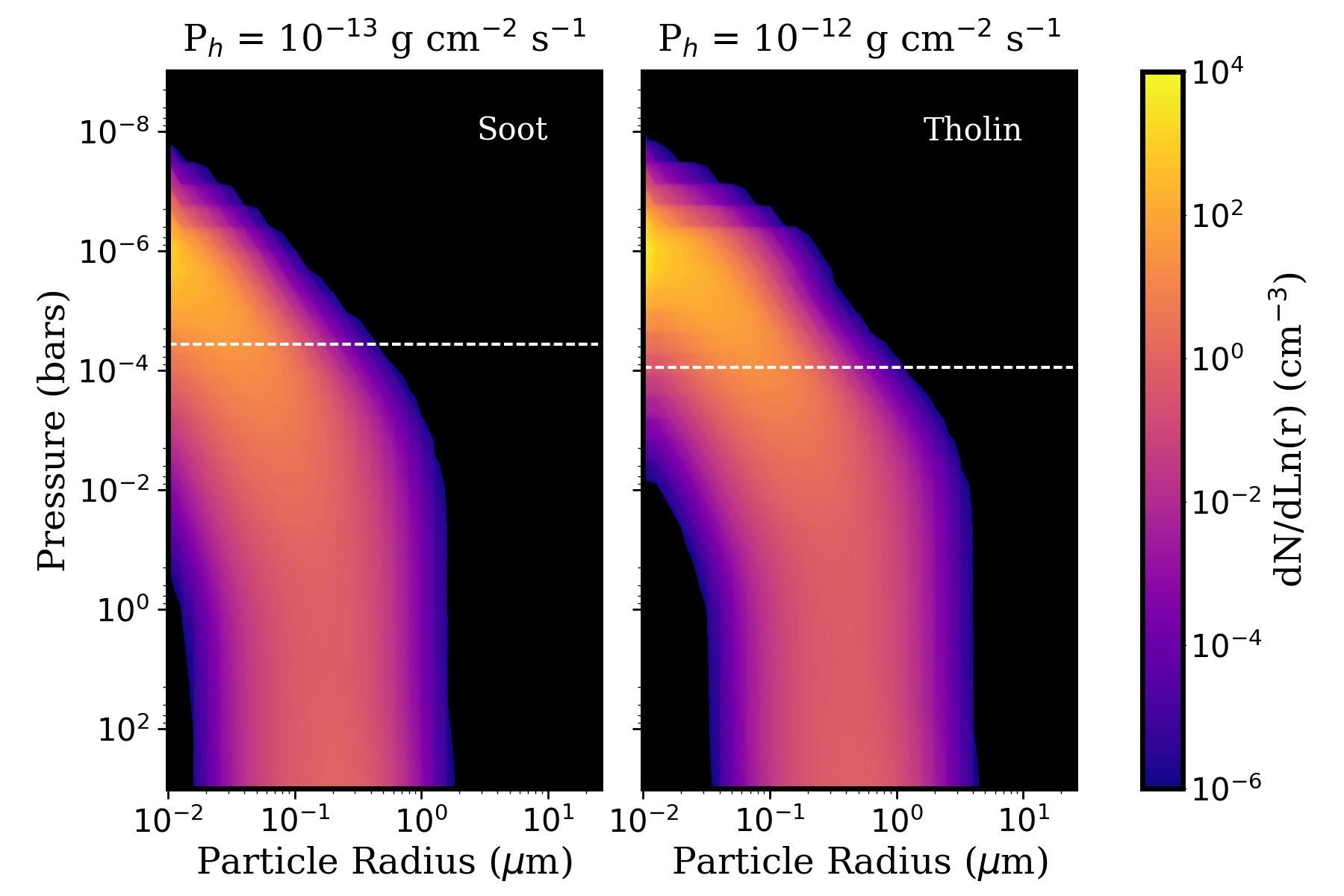}
\end{center}
\caption{Size and vertical distribution of haze particles for the best-fit soot and tholin haze models computed by CARMA. The colors indicate the number density of haze particles per logarithmic radius bin. The horizontal dashed white lines show the pressure levels where the optical depth in transmission at a wavelength of 1.5 $\mu$m is unity. In both cases, the hazes are dominated by submicron particles, with the smallest particle sizes in the upper atmosphere. The typical particle radii and opacity pressure levels are consistent with the values from our SCARLET retrieval.} \label{carmadist}
\end{figure}

The size and vertical distributions of the haze particles for the soot and tholin models are shown in Figure~\ref{carmadist}. The color coding indicates the number density of particles per logarithmic radius bin. For both cases, the haze distribution is dominated by submicron particles, particularly at the lowest pressure levels (below 0.1~mbar), consistent with the observed Rayleigh scattering slope in visible wavelengths. The horizontal dashed lines denote the highest pressure probed by our observations (i.e., optical depth of unity in transmission). Notably, the modeled particle size distributions and the opacity pressure levels are in agreement with the corresponding values $\log R_{\mathrm{part}}$ and $\log P_{\tau=1}$ inferred from the SCARLET retrieval (Table~\ref{retrievalresults}) to within the $1\sigma$ uncertainties. The demonstrated agreement between the retrieval and the CARMA results serves as an illustrative example of the increasing explanatory power of current aerosol models that incorporate detailed microphysical calculations and account for the opacity contributions from photochemical hazes.

\section{Constraints from secondary eclipse measurements}

The secondary eclipse measurements offer an independent look at the atmosphere of HAT-P-12b. While the transmission spectrum directly probes the day--night terminators, the secondary eclipse depths indicate the total outgoing flux from the dayside hemisphere relative to the star's flux. In Section~\ref{subsec:ecl}, we calculated depths of $0.042\%\pm 0.013\%$ and $0.045^{+0.017}_{-0.019}\%$ at 3.6 and 4.5~$\mu$m, respectively.

From these values, we can estimate the blackbody brightness temperature of the dayside hemisphere. We account for the uncertainties in the stellar parameters by deriving empirical analytical functions for the integrated stellar flux in the Spitzer bandpasses. This is done by fitting a polynomial in ($T_{\mathrm{eff}}$, [M/H], $\log g$) to the calculated stellar flux for a grid of \texttt{ATLAS} models \citep{atlas} spanning the ranges $T_{\mathrm{eff}}=[4000,5000]$~K, $[\mathrm{M/H}]=[-1.0,+0.5]$, and $\log g=[4.5,5.0]$. We then computed the posterior distribution of the dayside brightness temperature using a Monte Carlo sampling method, given priors on the stellar properties from \citet{hartman}.

We obtain brightness temperature estimates of $980^{+80}_{-100}$~K at 3.6~$\mu$m and $810^{+90}_{-160}$~K at 4.5~$\mu$m. We also find that both eclipse depths are consistent with a single blackbody temperature of $890^{+60}_{-70}$~K. This estimate is consistent at the $1.1\sigma$ level with the terminator temperature of $1010\pm80$~K previously derived from an analysis of the HST STIS transmission spectrum when assuming Rayleigh scattering \citep{singstis}. The predicted dayside equilibrium temperature of HAT-P-12b assuming zero albedo is 1150~K if incident energy is reradiated from the dayside only and 970~K if the planet reradiates the absorbed energy uniformly over the entire surface. The relatively low calculated dayside temperature indicates very efficient day--night recirculation of incident energy and possibly a nonzero albedo.

The Spitzer secondary eclipse depths can also provide constraints on atmospheric metallicity. Specifically, the ratio between the 3.6 and 4.5~$\mu$m depths varies systematically with metallicity. From the bottom left panel of Figure~\ref{spectrum}, we see the comparison between the measured depths and model spectra generated by SCARLET. The constraints provided by the Spitzer secondary eclipse depths in the combined transmission and emission spectra retrieval are weak due to the low signal-to-noise of the planetary flux detection as well as the low wavelength resolution of the two broadband points. Examining the model emission spectra, we can see diagnostic features in the 3--5~$\mu$m region that could be adequately probed with even modest wavelength resolution ($R\sim 20-30$). Near-future instruments, such as NIRSpec on the James Webb Space Telescope (JWST), will enable detailed studies of planetary emission spectra spanning the thermal infrared, opening up a new domain for exoplanet atmospheric characterization.

\section{Conclusions}
We have presented eight transit observations of the warm sub-Saturn HAT-P-12b obtained from HST and Spitzer. The resulting transmission spectrum from a joint analysis of all transit light curves covers the optical and near-infrared wavelength range from 0.3 to 5.0~$\mu$m. We obtain precise, updated estimates for the orbital parameters of the system. 

The main features of the transmission spectrum are a weak water vapor absorption feature at 1.4~$\mu$m and a prominent Rayleigh scattering slope throughout the visible wavelength range with no detected alkali absorption peaks. These features indicate significant cloud opacity in the atmosphere of HAT-P-12b, with a strong contribution from small-particle scattering in the upper atmosphere. The detection of Rayleigh scattering in the transmission spectrum and the low stellar activity of the host star make HAT-P-12b an important test case for evaluating the relationship between optical scattering slopes and stellar activity.

We have complemented our analysis of the transmission spectrum with new fits of secondary eclipse light curves in the 3.6 and 4.5~$\mu$m Spitzer bandpasses, from which we derive the depths $0.042\%\pm0.013\%$ and $0.045\%\pm0.018\%$, respectively. The dayside atmosphere is consistent with a single blackbody temperature of $890^{+60}_{-70}$~K and efficient day--night heat recirculation.

Through a multifaceted approach combining atmospheric retrievals from SCARLET using both transmission and emission spectra with the results of the aerosol microphysics model CARMA, we find that the atmosphere of HAT-P-12b has a near-solar C/O ratio of $0.48_{-0.37}^{+0.10}$ and an atmospheric metallicity that broadly spans the range between several tens and a few hundred times solar. While condensate cloud models produce particles that are too large to reproduce the observed Rayleigh scattering slope, models incorporating photochemical hazes consisting of tholins or soot readily generate submicron particles in the upper atmosphere and match the full range of observations. The aerosol modeling indicates moderate vertical mixing (eddy diffusion coefficient $K_{zz}= 10^{8}$~cm$^{2}$~s$^{-1}$) and opacity pressure levels around 0.1~mbar, consistent with the results of the retrievals.

HAT-P-12b fits within the growing population of well-characterized cooler exoplanets that show evidence for photochemical hazes. The temperature range spanned by these planets allows for the formation of an enormous diversity of condensate species \citep[e.g.,][]{singstis}. The importance of clouds and hazes in interpreting observed transmission spectra and their wide-ranging effects on atmospheric chemistry and dynamics illustrates the need for continued refinement in our understanding of the myriad physical and chemical processes that govern the formation and distribution of condensates in exoplanetary atmospheres. While current state-of-the-art cloud and haze models are becoming more sophisticated and capable of describing observations of individual exoplanets and observed trends in exoplanet cloudiness, there remain significant gaps in our knowledge of the detailed microphysics of \textit{ab initio} aerosol formation and the effects of secondary processes such as vertical mixing.

Our work also underscores the importance of increased spectral resolution in amplifying the explanatory power of both transmission and emission spectroscopy. The broadband Spitzer photometry at 3.6 and 4.5~$\mu$m has provided weak complementary constraints on the more discerning transmission spectra at shorter wavelengths. However, with even moderately increased spectral resolution in the 2--5~$\mu$m region, we can obtain much more precise estimates of atmospheric metallicity and C/O ratio and probe the absorption and emission features of a wide range of major atmospheric species. The capabilities of upcoming space-based telescopes such as JWST in this regard will usher in a new era of exoplanet atmospheric characterization.

\qquad

This work is based on observations with the NASA/ESA \textit{Hubble Space Telescope}, obtained at the Space Telescope Science Institute (STScI) operated by AURA, Inc. This work is also based in part on observations made with the \textit{Spitzer Space Telescope}, which is operated by the Jet Propulsion Laboratory, California Institute of Technology, under a contract with NASA. The research leading to these results has received funding from the European Research Council under the European Union's Seventh Framework Program (FP7/2007-2013)/ERC grant agreement No. 336792. Support for this work was also provided by NASA/STScI through grants linked to the HST-GO-12473 and HST-GO-14767 programs. I.W. and P.G. are supported by Heising-Simons Foundation \textit{51 Pegasi b} postdoctoral fellowships. H.A.K. acknowledges support from the Sloan Foundation.

\appendix

\section{HST Light Curves}

The following plots show the spectroscopic light curves for the six HST WFC3 and STIS transit observations. The left panel shows the light curves for each of the wavelength bins, corrected for instrumental systematics and arranged top to bottom in the order listed in Table~\ref{tab:fit2}. The best-fit transit light curves are overplotted in black. The right panel shows the corresponding residuals in parts per thousand (ppt). The error bars on all data points are set to the best-fit photometric noise parameter.

\begin{figure*}[h!]
\begin{center}
\includegraphics[width=17cm]{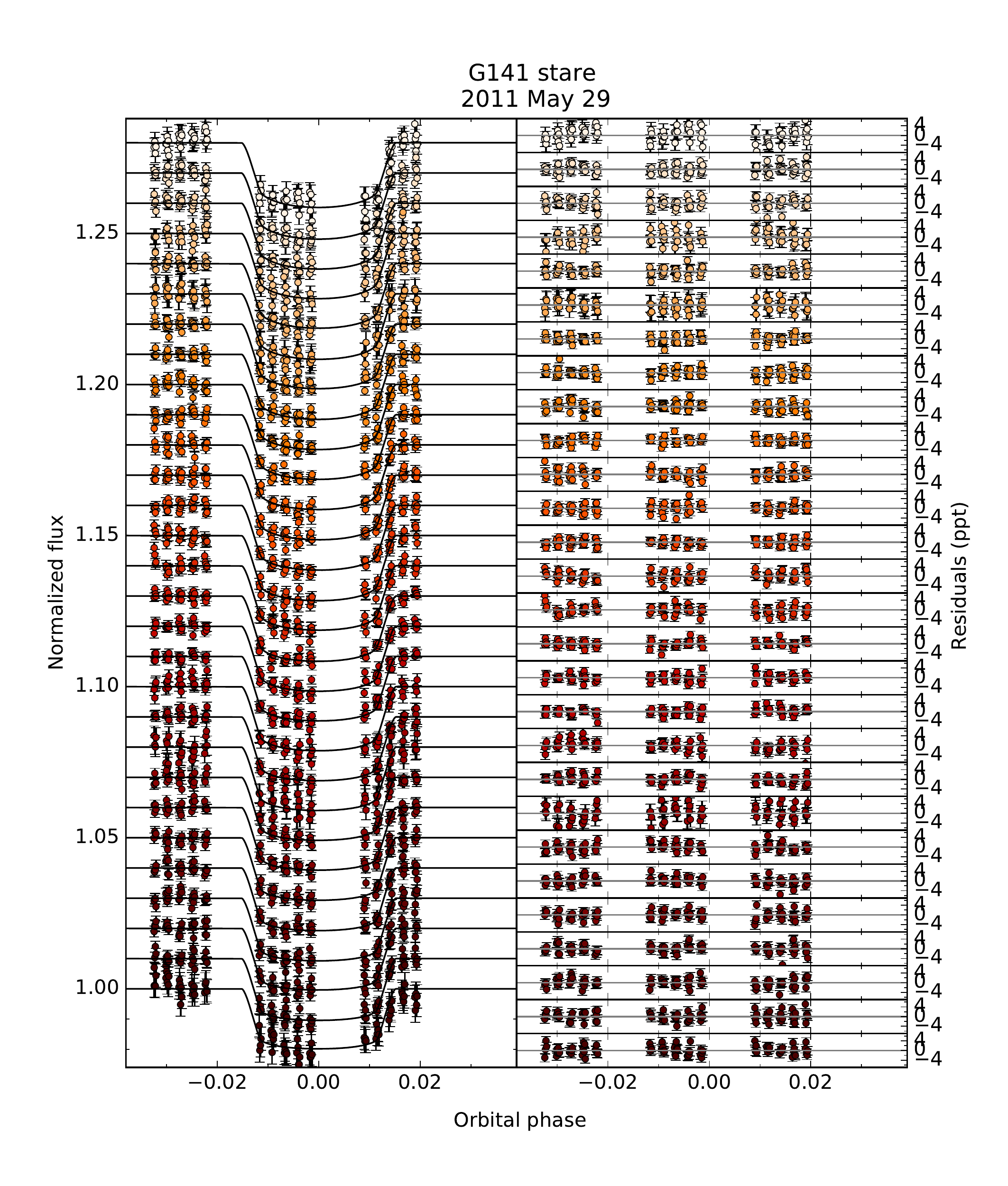}
\end{center}
\end{figure*}

\begin{figure*}[h!]
\begin{center}
\includegraphics[width=17cm]{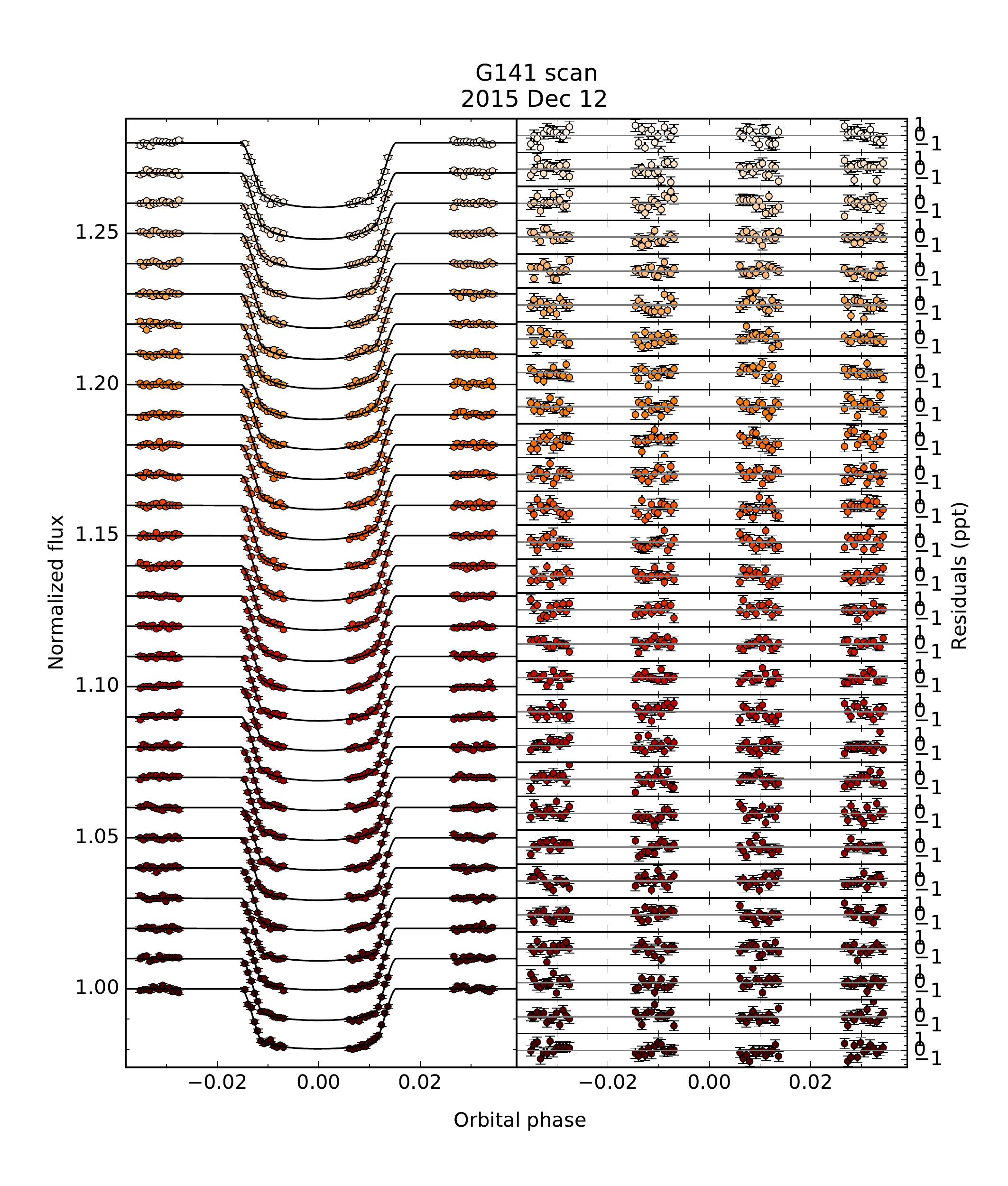}
\end{center}
\end{figure*}

\begin{figure*}[h!]
\begin{center}
\includegraphics[width=17cm]{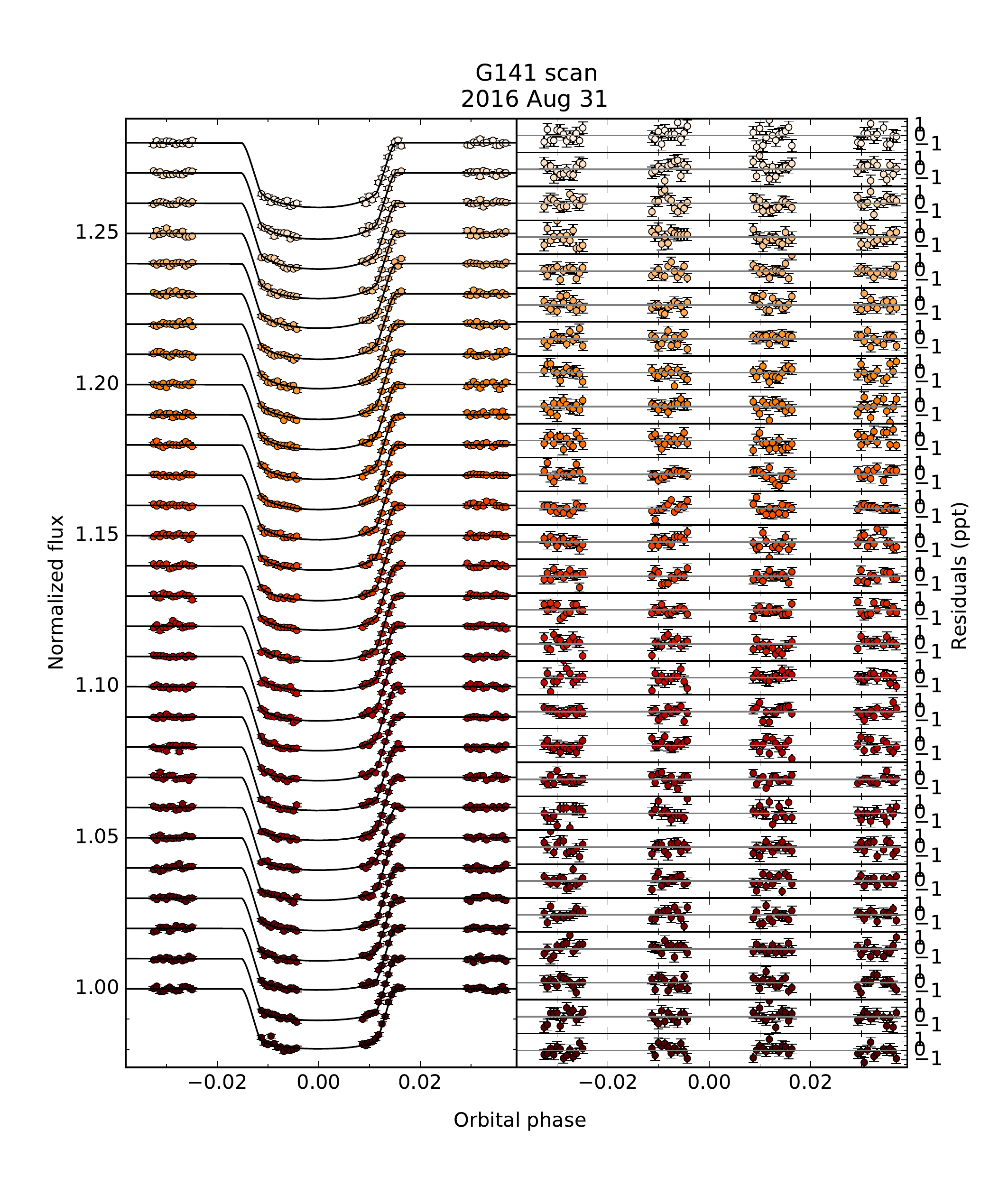}
\end{center}
\end{figure*}

\begin{figure*}[h!]
\begin{center}
\includegraphics[width=17cm]{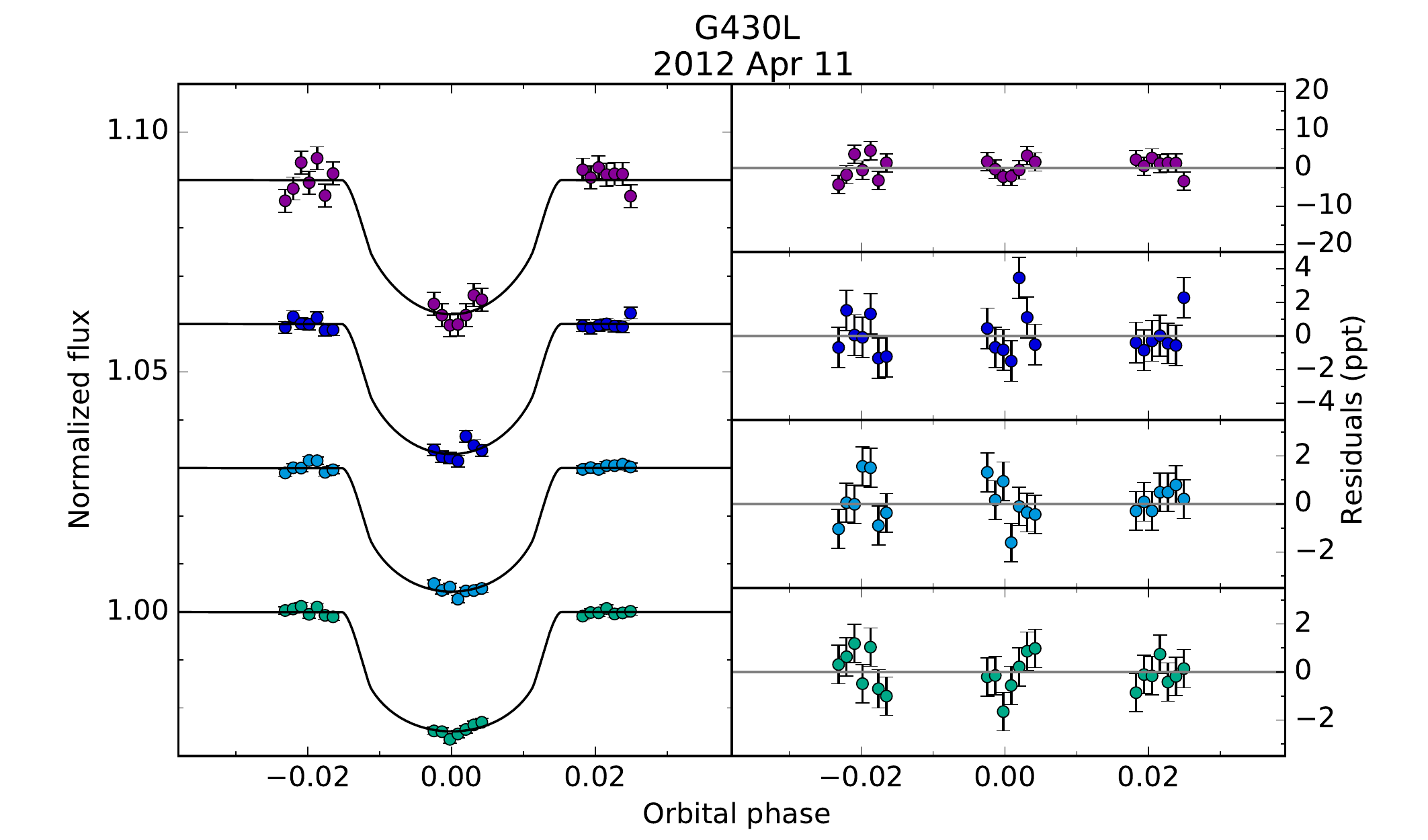}
\end{center}
\end{figure*}

\begin{figure*}[h!]
\begin{center}
\includegraphics[width=17cm]{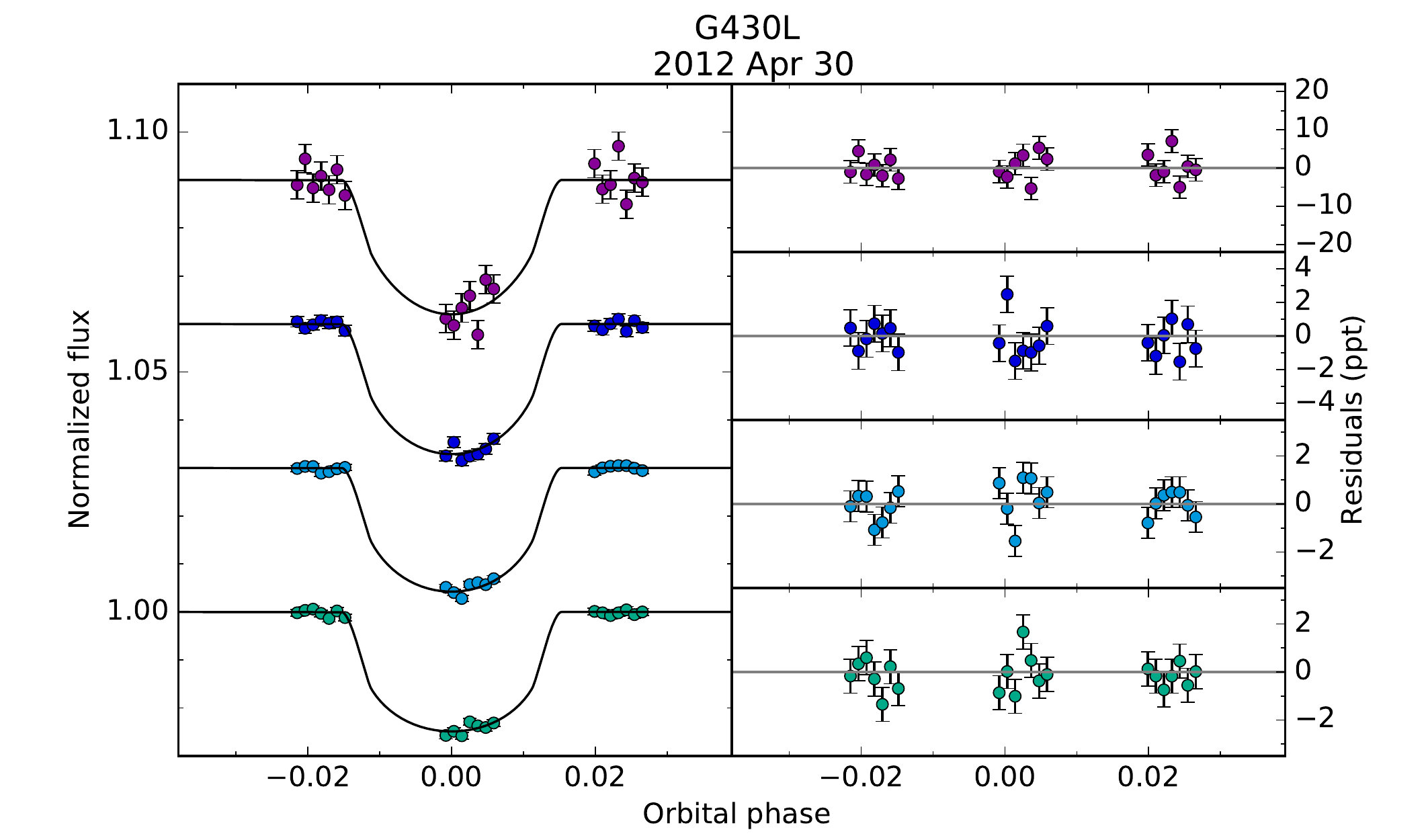}
\end{center}
\end{figure*}

\begin{figure*}[h!]
\begin{center}
\includegraphics[width=17cm]{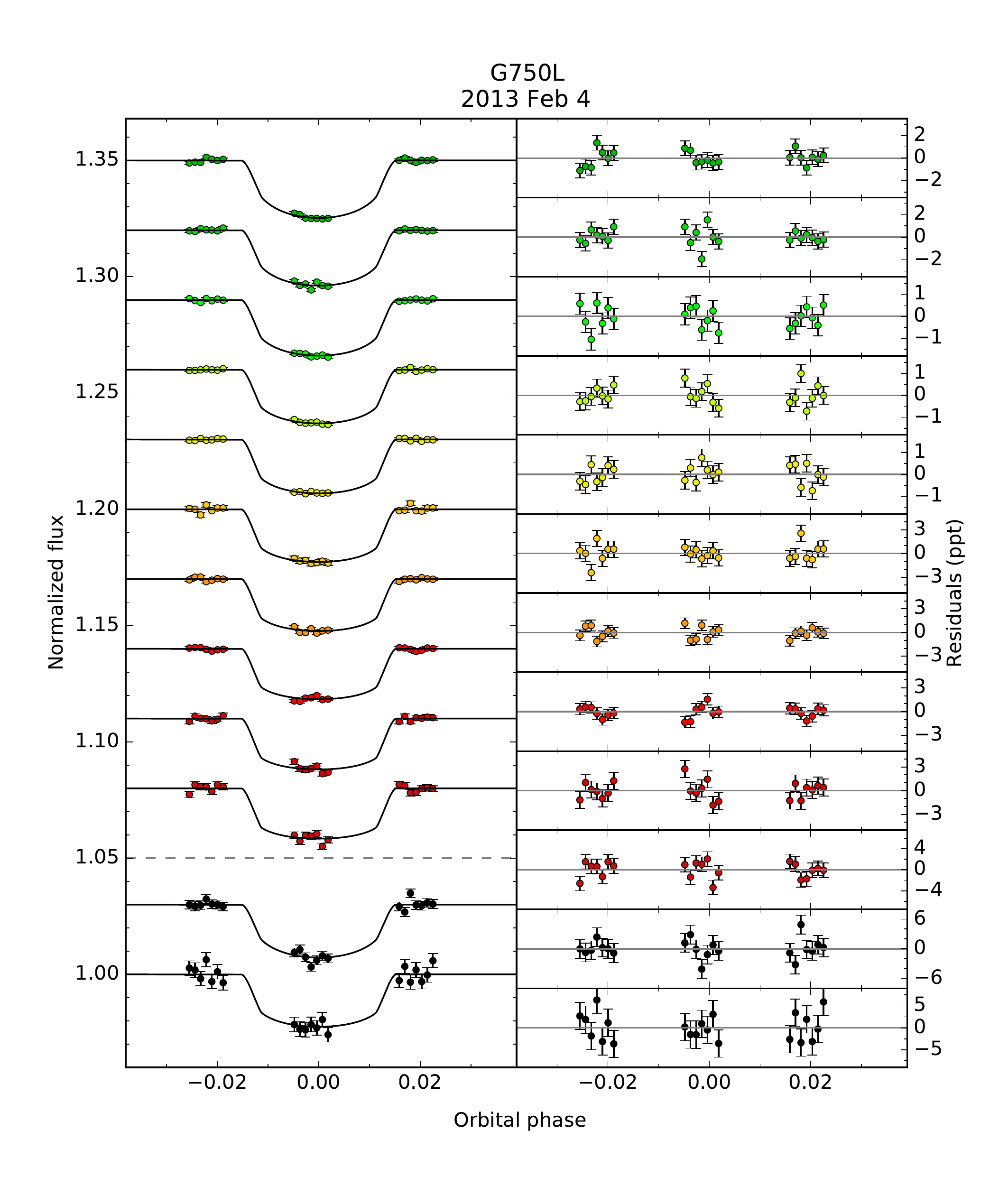}
\end{center}
\end{figure*}

\section{Cloud and haze modeling with CARMA}
CARMA was initially developed to investigate aerosol processes on Earth \citep{toon1979,turco}, and has since been adapted to various solar system bodies  \citep{toon1992,james,colaprete,barth2006,gao2014,gao2017} and exoplanets \citep{gao,powell2018,powell2019,adams2019}. Here we use the exoplanet version of CARMA, which has the ability to model clouds composed of a variety of species predicted by equilibrium chemistry and kinetic cloud formation models, including KCl, ZnS, Na$_{2}$S, MnS, Cr, Mg$_{2}$SiO$_{4}$, Fe, TiO$_{2}$, and Al$_{2}$O$_{3}$ \citep[see, for example, the review in][]{marley2013}.  We refer the reader to \citet{gao} and \citet{powell2019} for the relevant material properties of the condensates and the formation pathways we consider. Briefly, we consider homogeneous nucleation for species that can undergo direct phase change (TiO$_2$, Fe, Cr, and KCl) and heterogeneous nucleation for species that form via thermochemical reactions, represented in the gas phase by their limiting species (Al$_2$O$_3$: Al, Mg$_2$SiO$_4$: Mg, MnS: Mn, Na$_2$S: Na, ZnS: Zn; see, for example, \citealt{visscher2006}, \citealt{visscher2010}, and \citealt{morley2012}). Here TiO$_2$ is chosen to be the condensation nuclei of Al$_2$O$_3$, Mg$_2$SiO$_4$, MnS, and Na$_2$S due to its low energy barrier to homogeneous nucleation \citep[e.g.,][]{lee2018}, while KCl acts as the condensation nuclei to ZnS, as they both form at lower temperatures than the other condensates. Here Fe and Cr are also allowed to heterogeneously nucleate on TiO$_2$. The resulting cloud particles are either pure, in the case of the homogeneously nucleated particles, or a core surrounded by a mantle, in the case of the heterogeneously nucleated particles. This is a simplification of the mixed-grains formalism of other kinetic cloud models 
\citep[e.g.,][]{helling2016,lee2016,lines2}. Cloud particles of different compositions do not interact, and their size distributions are computed independently of each other, except in the case of condensation nuclei and mantling species; i.e. formation of the latter depletes the former.

Each cloud simulation begins with a background H$_{2}$/He atmosphere devoid of cloud particles, with condensate vapor only at the deepest atmospheric level. We use GGchem \citep{ggchem} to set the initial mixing ratio of each condensate species at this lower boundary. As the simulation advances, all condensate vapors are mixed upward via eddy diffusion and parameterized by the eddy diffusion coefficient $K_{zz}$, until they either become well mixed in the atmosphere or achieve supersaturation. Particle nucleation and condensation may then occur, provided that the supersaturation is sufficiently large to overcome the nucleation energy barriers of the various condensate species. Cloud particle formation depletes the condensate vapors until their resupply by eddy diffusion from depth is sufficient to balance. We do not explicitly consider any gas chemistry in our modeling. Growth of cloud particles by coagulation and vertical transport of cloud particles proceed until a steady state is reached.

CARMA also models coagulation and vertical transport of photochemical hazes, following the methodology developed in \citet{gao2017} and \citet{adams2019}. The detailed chemical pathways and formation efficiencies of exoplanet hazes are much more complex and less understood than those predicted for condensation clouds \citep{fleury2018,he2018,horst}, and CARMA does not explicitly model the production of aerosol particles. Instead, we choose to model haze production generically by setting the haze production rate as a free parameter. We assume spherical haze particles with a mass density of 1~g~cm$^{-3}$ and a minimum radius of 10~nm; it has been shown that the minimum particle radius does not strongly affect the optical depth at equilibrium \citep[e.g.,][]{adams2019}. We do not consider condensation when modeling hazes. 

Haze simulations also begin with a background H$_{2}$/He atmosphere devoid of aerosols. As the simulation advances, 10~nm haze particles are produced at high altitudes, after which they can grow by coagulation and are transported into the deep atmosphere by sedimentation and eddy diffusion. Haze particles are assumed to evaporate at the lower boundary of the model, though it does not impact the resulting transmission spectra, since the lower boundary is set at pressures $>$10 bar.

To generate predicted transmission spectra from the forward-modeled aerosol distributions, we first use the \texttt{pymiecoated} tool to compute the extinction efficiency, single scattering albedo, and asymmetry factor of the aerosol particles. The refractive indices for the various aerosol species are compiled from \citet{posch2003}, \citet{zeidler2011}, \citet{morley2012}, \citet{wakefordsing}, and \citet{lavvas2017}. We then use a standard 1D radiative transfer model to produce the transmission spectra \citep[e.g.,][]{fortney2010}. These spectra are subsequently binned to the resolution of the observations, and the base planet radius is shifted to best fit the observed transmission spectrum. We compute the RCS goodness-of-fit metric to choose the best-performing model runs.

\quad

\end{document}